\documentclass[zpreprint,zbstdefault]{./LaTeX/zeus/zeus_paper}
\usepackage[english]{babel}

\newcommand{\ZcoosysB}{%
The ZEUS coordinate system is a right-handed Cartesian system, with the $Z$
axis pointing in the proton beam direction, referred to as the ``forward
direction'', and the $X$ axis pointing left towards the centre of HERA.
The coordinate origin is at the nominal interaction point.\xspace}
\newcommand{\Zpsrap}{%
The pseudorapidity is defined as $\eta=-\ln\left(\tan\frac{\theta}{2}\right)$,
where the polar angle, $\theta$, is measured with respect to the proton beam
direction.\xspace}

\newcommand{\ZcoosysfnBeta}{\footnote{\ZcoosysB\Zpsrap}}
\newcommand{\Zdetdesc}{%
A detailed description of the ZEUS detector can be found 
elsewhere~\cite{zeus:1993:bluebook}. A brief outline of the 
components that are most relevant for this analysis is given
below.\xspace}
\newcommand{\Zctddesc}[1]{%
Charged particles are tracked in the central tracking detector (CTD)~\citeCTD,
which operates in a magnetic field of $1.43\Tesla$ provided by a thin 
superconducting coil. The CTD consists of 72~cylindrical drift chamber 
layers, organized in 9~superlayers covering the polar-angle#1 region 
\mbox{$15^\circ<\theta<164^\circ$}. The transverse-momentum resolution for
full-length tracks is $\sigma(p_T)/p_T=0.0058p_T\oplus0.0065\oplus0.0014/p_T$,
with $p_T$ in $\Gev$.}
\newcommand{\Zcaldesc}{%
The high-resolution uranium--scintillator calorimeter (CAL)~\citeCAL consists 
of three parts: the forward (FCAL), the barrel (BCAL) and the rear (RCAL)
calorimeters. Each part is subdivided transversely into towers and
longitudinally into one electromagnetic section (EMC) and either one (in RCAL)
or two (in BCAL and FCAL) hadronic sections (HAC). The smallest subdivision of
the calorimeter is called a cell.  The CAL energy resolutions, as measured under
test-beam conditions, are $\sigma(E)/E=0.18/\sqrt{E}$ for electrons and
$\sigma(E)/E=0.35/\sqrt{E}$ for hadrons ($E$ in $\Gev$).}

\chardef\usc=95
\chardef\til=126
\catcode`\@=11 
\DeclareRobustCommand\xdotspace{\futurelet\@let@token\@xdotspace}
\def\@xdotspace{%
  \ifx\@let@token.\else
  \ifx\@let@token\bgroup.\else
  \ifx\@let@token\egroup.\else
  \ifx\@let@token\/.\else
  \ifx\@let@token\ .\else
  \ifx\@let@token~.\else
  \ifx\@let@token!.\else
  \ifx\@let@token,.\else
  \ifx\@let@token:.\else
  \ifx\@let@token;.\else
  \ifx\@let@token?.\else
  \ifx\@let@token/.\else
  \ifx\@let@token'.\else
  \ifx\@let@token).\else
  \ifx\@let@token-.\else
  \ifx\@let@token\@xobeysp.\else
  \ifx\@let@token\space.\else
  \ifx\@let@token\@sptoken.\else
   .\space
   \fi\fi\fi\fi\fi\fi\fi\fi\fi\fi\fi\fi\fi\fi\fi\fi\fi\fi}
\catcode`\@=12 

\newcommand{\stru}[2]{%
   \relax\ifmmode\hbox{\vrule height#1 depth#2 width0pt}%
   \else\vrule height#1 depth#2 width0pt\fi}

\newcommand{\Ronum}[1]{\uppercase\expandafter{\romannumeral#1}}
\newcommand{\ronum}[1]{\expandafter{\romannumeral#1}}
\DeclareRobustCommand{\LaTeXZ}{%
  \LaTeX\kern-.05em4\kern-.1em
  {\raisebox{-0.2ex}{$\scriptstyle\text{ZEUS}$}}\xspace}

\newcommand{\eq}[1]{(\ref{eq-#1})}

\newcommand{\fig}[1]{Fig.~\ref{fig-#1}}
\newcommand{\Fig}[1]{Figure~\ref{fig-#1}}
\newcommand{\figand}[2]{Figs.~\ref{fig-#1} and~\ref{fig-#2}}

\newcommand{\taband}[2]{Tables~\ref{tab-#1} and~\ref{tab-#2}}

\newcommand{\Sect}[1]{Section~\ref{sec-#1}}

\DeclareMathAlphabet{\mathbf}{OT1}{cmr}{bx}{sl}
\newcommand{\eVdist}{\kern-0.06667em}

\newcommand{\Gev}{{\text{Ge}\eVdist\text{V\/}}}

\newcommand{\mev}{{\,\text{Me}\eVdist\text{V\/}}}
\newcommand{\gev}{{\,\text{Ge}\eVdist\text{V\/}}}

\newcommand{\pbi}{\,\text{pb}^{-1}}

\newcommand{\Tesla}{\,\text{T}}

\newcommand{\slashfrac}[2]{%
  \raisebox{0.5ex}{\ensuremath #1}\kern-0.12em/\kern-0.08em
  \raisebox{-.8ex}{\ensuremath #2}}

\newcommand{\sqr}[3]{%
    {\vcenter{\hrule height.#3ex\hbox{\vrule width.#2ex height#1ex
     \kern#1ex\vrule width.#3ex}\hrule height.#2ex}}}

\newcommand{\widebar}[1]{%
   \mkern1.5mu\overline{\mkern-1.5mu#1\mkern-1.mu}\mkern1.mu}
\catcode`\@=11 
\newcommand{\parenbar}{\mathpalette\p@renb@r}
\def\p@renb@r#1#2{\vbox{%
  \ifx#1\scriptscriptstyle \dimen@.7em\dimen@ii.2em\else
  \ifx#1\scriptstyle \dimen@.8em\dimen@ii.25em\else
  \dimen@1em\dimen@ii.4em\fi\fi \offinterlineskip
  \ialign{\hfill##\hfill\cr
    \vbox{\hrule width\dimen@ii}\cr
    \noalign{\vskip-.3ex}%
    \hbox to\dimen@{$\mathchar300\hfil\mathchar301$}\cr
    \noalign{\vskip-.3ex}%
    $#1#2$\cr}}}
\catcode`\@=12 

\newcommand{\qbar}{\widebar{q}}

\newcommand{\als}{\alpha_s}

\newcommand{\diff}{{\rm d}}

\newcommand{\IP}{{\rm I$\kern-0.01667em$P}\xspace}

\mathchardef\qsm=63
\mathchardef\pls=43
\mathchardef\mns=512
\mathchardef\plm=518
\mathchardef\eql=61
\mathchardef\smallleft=300
\mathchardef\smallright=301
\mathchardef\les=316
\mathchardef\gre=318
\mathchardef\leq=532
\mathchardef\grq=533

\catcode`\@=11 
\newcounter{pict@width}
\newcounter{pict@height}
\newlength{\pict@scale}
\newcommand{\psfigadd}[4]{%
\setcounter{pict@width}{1*\ratio{#2+\pict@scale/2}{\pict@scale}}
\setcounter{pict@height}{1*\ratio{#3+\pict@scale/2}{\pict@scale}}
\setlength{\unitlength}{\pict@scale}
\hbox to #2{\hspace{-\fill}\begin{picture}(\thepict@width,\thepict@height)
\put(0,0){\psfig{figure=#1,width=#2,height=#3,clip=}}
\SetScale{0.283466457}
\SetWidth{1.763889}
{#4}
\end{picture}}
}
\newcounter{pict@widthfst}
\newcounter{pict@widthscd}
\newcounter{pict@widthtot}
\newcommand{\psfigaddtwo}[7]{%
\setcounter{pict@widthfst}{1*\ratio{#2+\pict@scale/2}{\pict@scale}}
\setcounter{pict@widthscd}{1*\ratio{#2+#4+\pict@scale/2}{\pict@scale}}
\setcounter{pict@widthtot}{1*\ratio{#2+#4+#6+\pict@scale/2}{\pict@scale}}
\setcounter{pict@height}{1*\ratio{#3+\pict@scale/2}{\pict@scale}}
\setlength{\unitlength}{\pict@scale}
\hbox{\hspace{-\fill}\begin{picture}(\thepict@widthtot,\thepict@height)
\put(0,0){\psfig{figure=#1,width=#2,height=#3,clip=}}
\put(\thepict@widthscd,0){\psfig{figure=#5,width=#6,height=#3,clip=}}
\SetScale{0.283466457}
\SetWidth{1.763889}
{#7}
\end{picture}}
}
\newcommand{\psfigror}[4]{%
\setcounter{pict@width}{1*\ratio{#2+\pict@scale/2}{\pict@scale}}
\setcounter{pict@height}{1*\ratio{#3+\pict@scale/2}{\pict@scale}}
\setlength{\unitlength}{\pict@scale}
\hbox{\begin{picture}(\thepict@width,\thepict@height)
\put(0,\thepict@height){\psfig{figure=#1,width=#3,height=#2,clip=,angle=270}}
\SetScale{0.283466457}
\SetWidth{1.763889}
{#4}
\end{picture}}
}
\newcommand{\psfigrol}[4]{%
\setcounter{pict@width}{1*\ratio{#2+\pict@scale/2}{\pict@scale}}
\setcounter{pict@height}{1*\ratio{#3+\pict@scale/2}{\pict@scale}}
\setlength{\unitlength}{\pict@scale}
\hbox{\begin{picture}(\thepict@width,\thepict@height)
\put(0,0){\psfig{figure=#1,width=#3,height=#2,clip=,angle=90}}
\SetScale{0.283466457}
\SetWidth{1.763889}
{#4}
\end{picture}}
}
\catcode`\@=12 
\newlength\listtextwidth

\catcode`\@=11 
\newlength{\@tabfninsert}
\newlength{\@tabfnwidth}
\newcommand{\tabfootnote}[2]{%
  \setlength{\@tabfninsert}{0.8em}
  \setlength{\@tabfnwidth}{\textwidth}
  \addtolength{\@tabfnwidth}{-\@tabfninsert}
  \addtolength{\@tabfnwidth}{-0.4em}
  \noindent\makebox[\@tabfninsert][r]{\footnotesize$^{#1}$\hfil}\hfill%
  \parbox[t]{\@tabfnwidth}{\footnotesize #2\hfill}}
\catcode`\@=12 

\def\citeCTD{{\cite{%
nim:a279:290,*npps:b32:181,*nim:a338:254%
}}\xspace}
\def\citeCAL{{\cite{%
nim:a309:77,*nim:a309:101,*nim:a321:356,*nim:a336:23%
}}\xspace}

\def\citeVDM{{\cite{%
anphy:11:1,*prl:22:981%
}}\xspace}

\def\citeREGGE{{\cite{%
collins:1977:regge%
}}\xspace}

\def\citeVM_HERA{{\cite{%
crittenden:1997:mesons,*rmp:71:1275%
}}\xspace}
 
\def\citeHIGHT95{{\cite{%
epj:c14:213%
}}\xspace}

\def\citeEPSOFTmk{{\cite{%
thesis:kasprzak:1994%
}}\xspace}

\def\citeEPSOFTla{{\cite{%
thesis:adamczyk:1999%
}}\xspace}
 
\def\citeDIFFVM{{\cite{%
proc:mc:1998:396%
}}\xspace}

\begin{document}
\prepnum{DESY--02--072} 

\title{
Measurement of proton-dissociative\\
diffractive photoproduction of vector mesons\\
at large momentum transfer at HERA
}                                                       
                    
\author{ZEUS Collaboration}
\date{May 2002}

\abstract{
Diffractive photoproduction of vector mesons, 
$\gamma p \to V Y$, where $Y$ is a proton-dissociative system, 
has been measured in $e^+p$ interactions with the ZEUS detector 
at HERA using an integrated luminosity of 25$\pbi$.
The differential cross section, $\diff\sigma/\diff t$, 
is presented for $-t<12\gev^2$, where $t$ is the square 
of the four-momentum transferred to the vector meson.
The data span the range in photon-proton centre-of-mass energy, $W$, 
from $80\gev$  to $120\gev$. 
The $t$ distributions are well fit by a power law, 
$\diff\sigma/\diff t \propto (-t)^{-n}$.
The slope of the Pomeron trajectory,
measured from the $W$ dependence of the $\rho^0$ and $\phi$
cross sections in bins of $t$, is consistent with zero.
The ratios 
$\diff\sigma_{\gamma p \to \phi Y}/\diff t$ to
$\diff\sigma_{\gamma p \to \rho^0 Y}/\diff t$
and
$\diff\sigma_{\gamma p \to J/\psi Y}/\diff t$ to
$\diff\sigma_{\gamma p \to \rho^0 Y}/\diff t$
increase with increasing $-t$. 
Decay-angle analyses for $\rho^0$, $\phi$ and $J/\psi$
mesons have been carried out.
For the $\rho^0$ and $\phi$ mesons, contributions 
from single and double helicity
flip are observed. The results are compared to expectations
of theoretical models.
}

\makezeustitle

\pagenumbering{Roman}                                                                              
\begin{center}                                                                                     
{                      \Large  The ZEUS Collaboration              }                               
\end{center}                                                                                       
  S.~Chekanov,                                                                                     
  D.~Krakauer,                                                                                     
  S.~Magill,                                                                                       
  B.~Musgrave,                                                                                     
  J.~Repond,                                                                                       
  R.~Yoshida\\                                                                                     
 {\it Argonne National Laboratory, Argonne, Illinois 60439-4815}~$^{n}$                            
\par \filbreak                                                                                     
  M.C.K.~Mattingly \\                                                                              
 {\it Andrews University, Berrien Springs, Michigan 49104-0380}                                    
\par \filbreak                                                                                     
  P.~Antonioli,                                                                                    
  G.~Bari,                                                                                         
  M.~Basile,                                                                                       
  L.~Bellagamba,                                                                                   
  D.~Boscherini,                                                                                   
  A.~Bruni,                                                                                        
  G.~Bruni,                                                                                        
  G.~Cara~Romeo,                                                                                   
  L.~Cifarelli,                                                                                    
  F.~Cindolo,                                                                                      
  A.~Contin,                                                                                       
  M.~Corradi,                                                                                      
  S.~De~Pasquale,                                                                                  
  P.~Giusti,                                                                                       
  G.~Iacobucci,                                                                                    
  G.~Levi,                                                                                         
  A.~Margotti,                                                                                     
  R.~Nania,                                                                                        
  F.~Palmonari,                                                                                    
  A.~Pesci,                                                                                        
  G.~Sartorelli,                                                                                   
  A.~Zichichi  \\                                                                                  
  {\it University and INFN Bologna, Bologna, Italy}~$^{e}$
% ...
\newpage                                                                                           
% ...
                                                           %
                                                           %
\newpage   
                                                           %
                                                           %
\begin{tabular}[h]{rp{14cm}}                                                                       
$^{a}$ &  supported by the Natural Sciences and Engineering Research                               
          Council of Canada (NSERC) \\                                                             
$^{b}$ &  supported by the German Federal Ministry for Education and                               
          Research (BMBF), under contract numbers HZ1GUA 2, HZ1GUB 0, HZ1PDA 5, HZ1VFA 5\\         
$^{c}$ &  supported by the MINERVA Gesellschaft f\"ur Forschung GmbH, the                          
          Israel Science Foundation, the U.S.-Israel Binational Science                            
          Foundation, the Israel Ministry of Science and the Benozyio Center                       
          for High Energy Physics\\                                                                
$^{d}$ &  supported by the German-Israeli Foundation, the Israel Science                           
          Foundation, and by the Israel Ministry of Science\\                                      
$^{e}$ &  supported by the Italian National Institute for Nuclear Physics (INFN) \\                
$^{f}$ &  supported by the Japanese Ministry of Education, Science and                             
          Culture (the Monbusho) and its grants for Scientific Research\\                          
$^{g}$ &  supported by the Korean Ministry of Education and Korea Science                          
          and Engineering Foundation\\                                                             
$^{h}$ &  supported by the Netherlands Foundation for Research on Matter (FOM)\\                   
$^{i}$ &  supported by the Polish State Committee for Scientific Research,                         
          grant no. 620/E-77/SPUB-M/DESY/P-03/DZ 247/2000-2002\\                                   
$^{j}$ &  partially supported by the German Federal Ministry for Education                         
          and Research (BMBF)\\                                                                    
$^{k}$ &  supported by the Fund for Fundamental Research of Russian Ministry                       
          for Science and Edu\-cation and by the German Federal Ministry for                       
          Education and Research (BMBF)\\                                                          
$^{l}$ &  supported by the Spanish Ministry of Education and Science                               
          through funds provided by CICYT\\                                                        
$^{m}$ &  supported by the Particle Physics and Astronomy Research Council, UK\\                   
$^{n}$ &  supported by the US Department of Energy\\                                               
$^{o}$ &  supported by the US National Science Foundation\\                                        
$^{p}$ &  supported by the Polish State Committee for Scientific Research,                         
          grant no. 112/E-356/SPUB-M/DESY/P-03/DZ 301/2000-2002, 2 P03B 13922\\                    
$^{q}$ &  supported by the Polish State Committee for Scientific Research,                         
          grant no. 115/E-343/SPUB-M/DESY/P-03/DZ 121/2001-2002, 2 P03B 07022\\                    
\end{tabular}                                                                                      
                                                           %
                                                           %
\clearpage   

\pagenumbering{arabic} 
\pagestyle{plain}
\section{Introduction}
\label{sec-int}
Studies of the elastic production of vector mesons in electron-proton
interactions~\citeVM_HERA show that at large $Q^2$, 
the exchanged-photon virtuality, or at high vector-meson mass, 
the cross section increases with energy faster than is
observed in hadron-hadron 
interactions~\cite{np:b231:189,pl:b395:311}. The latter increase 
is well described by Regge theory~\citeREGGE. 
Although the energy dependence of vector-meson production 
in $ep$ interactions 
can be described by models based on Regge 
phenomenology~\cite{pl:b470:243,*pl:b478:146,*hep-ph-0112242}
and the vector dominance model (VDM)~\citeVDM,
it can also be explained by models based on 
perturbative QCD (pQCD)~\cite{zfp:c57:89,pr:d50:3134}. In this case, 
it is related to the rise of the gluon density in the proton 
as $x$ decreases, where $x$ is the Bjorken scaling variable.
In pQCD models, a perturbative (hard) scale can be provided by either
a high $Q^2$ or a large meson mass.  
It is also predicted~\cite{prl:63:1914,*pl:b284:123} 
that, in diffractive dissociation of hadrons, the squared 
four-momentum transfer, $t$, may serve as a hard scale.
 
The subject of this paper is vector-meson ($V$) photoproduction 
at high $-t$, which is dominated
by the proton-dissociative reaction, $\gamma p \to V Y$, 
where $Y$ is the dissociated hadronic system.
The data cover the photon-proton centre-of-mass energy range 
\mbox{$80<W<120$} $\gev$ 
and extend from $-t=1.1  \gev^2$ up to $-t = 12 \gev^2$  
for the $\rho^0$, to $7 \gev^2$ for the $\phi$ and 
to $6.5 \gev^2$ for the $J/\psi$ meson. 
The vector mesons were identified 
via their decays to two oppositely charged particles: 
$\rho^0 \to \pi^+\pi^-$, $\phi \to K^+K^-$, and 
$J/\psi \to l^+l^-(e^+e^-,\mu^+\mu^-)$. 
%
%
\section{Theoretical approaches}

\subsection{The Regge model}
In Regge theory, diffractive processes are assumed to proceed through
the exchange of the Pomeron trajectory.
The vector dominance model, schematically indicated in~\fig{feynman}(a),  
together with Regge theory, gives the following form for the 
double-differential cross section for the reaction $\gamma p \to VY$:
\begin{equation}
\frac{\diff^2\sigma_{\gamma p \to V Y}}{\diff M_Y^2 \diff t} =
f(t)\frac{1}{W^2}\left(\frac{W^2}{M_Y^2}\right)^{2\alpha(t)-1}(M_Y^2)^{\alpha(0)-1}, 
\label{eq-vdm}
\end{equation} 
where $M_Y$ is the mass of the diffractively produced hadronic state, $Y$.

Assuming a linear form for the Pomeron trajectory, 
$\alpha(t) = \alpha(0) + \alpha^\prime t$,
fits~\cite{epj:c14:213,hep-ex-0201043} to the elastic
photoproduction of $\rho^0$, $\phi$ and $J/\psi$ mesons in the range 
$-t < 1.5 \gev^2$ gave 
$\alpha(0) = 1.096 \pm  0.021$ and  $\alpha^\prime = 0.125 \pm 0.038\gev^{-2}
$ for the $\rho^0$, $\alpha(0) = 1.081 \pm  0.010$ and  
$\alpha^\prime = 0.158 \pm 0.028\gev^{-2}$ for the $\phi$, and
$\alpha(0) = 1.200 \pm  0.009 ^{+0.004}_{-0.010}$ and 
$\alpha^\prime = 0.115 \pm 0.018^{+0.008}_{-0.015} \gev^{-2}$ for 
the $J/\psi$ meson.
These values may be compared to $\alpha(0)=1.08$~\cite{np:b231:189},
$\alpha(0)=1.096$~\cite{pl:b395:311} and 
$\alpha^\prime= 0.25 \gev^{-2}$~\cite{np:b231:189}, obtained from 
fits to hadron-scattering data at $-t < 0.5 \gev^2$. 
These numbers indicate that the Pomeron trajectory is not universal.
In fact, it has been   
suggested~\cite{npps:99a:24} that, in the production of vector mesons,
an anomalous Regge trajectory exchange, with $\alpha(0) \approx 1$ and
$\alpha^\prime \approx 0 $, gives a dominant contribution
to the differential cross section at large $-t$.   

\subsection{pQCD models}
\label{sec-pqcd}
\begin{figure}[p]
\vfill
\begin{center}
\epsfig{file=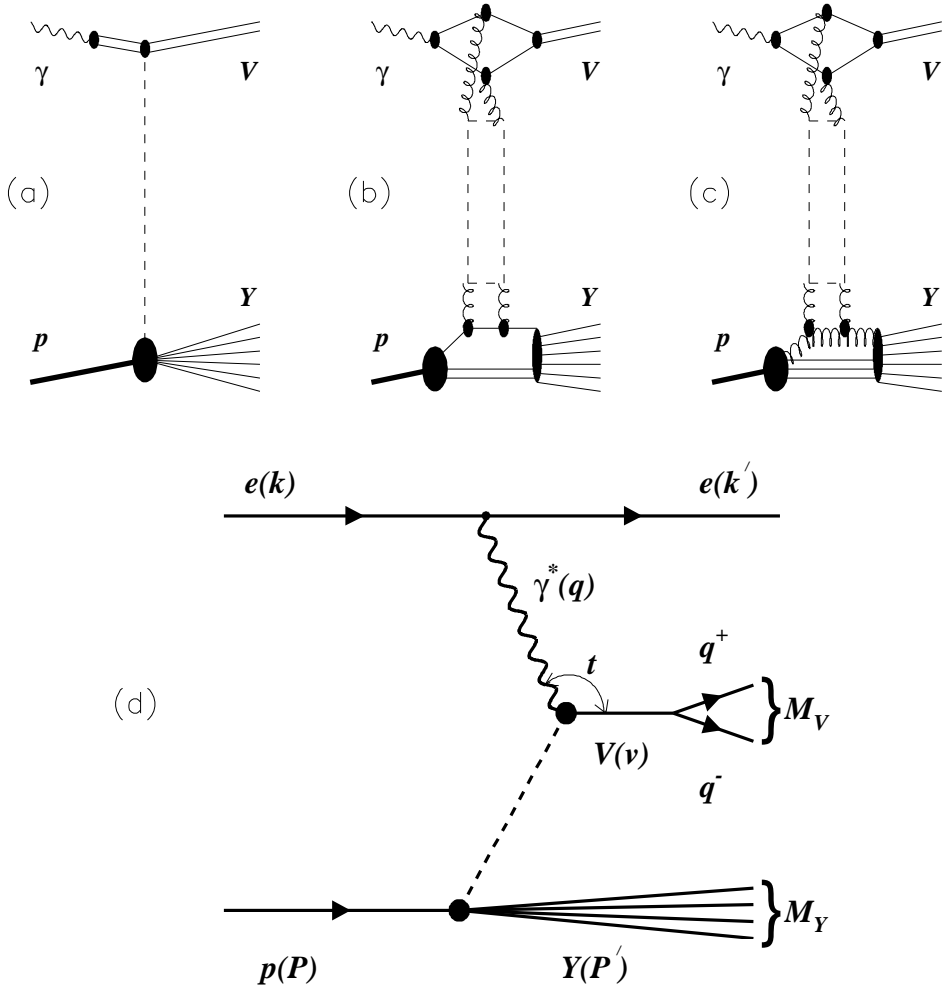,
width=1.0\textwidth}
\end{center}
\caption{(a) $\gamma p \to V Y$ scattering in the VDM model. The photon
fluctuates into a vector meson, which then scatters off the proton and the 
proton
breaks up. (b,c) Example of $\gamma p \to V Y$ scattering 
in pQCD models. The photon
fluctuates into a $q\qbar$ pair, which then scatters off a single  
quark (b) or gluon (c) in the proton by the exchange of  two gluons or a
gluon ladder. The scattered $q\qbar$ pair
becomes a vector meson. The struck parton and the proton remnant 
fragment into a system of hadrons. (d) Schematic diagram 
of proton-dissociative vector-meson production
in $ep$ interactions, $e p \to e V Y$. 
}
  \label{fig-feynman}
\vfill
\end{figure}

In models based on pQCD
~\cite{zfp:c68:137,pr:d53:3564,pr:d54:5523,pl:b375:301,pl:b478:101,hep-ph-0107068}, the reaction $\gamma p \to VY$ 
is viewed as a sequence of three
successive processes, illustrated in Figs.~\ref{fig-feynman}(b) 
and~\ref{fig-feynman}(c): the photon
fluctuates into a $q\qbar$ pair; the $q\qbar$ pair scatters off a single 
parton in the proton by the exchange of a colour singlet; 
the scattered $q\qbar$ pair
becomes a vector meson and the struck parton and the proton remnant 
fragment into a system of hadrons. The  probability of the photon
fluctuating into a $q\qbar$ pair is 
parameterised 
by the photon wave-function. Many models assume that 
the interaction 
of the $q\qbar$ pair with a parton in the proton is mediated in the
lowest order by the exchange of two 
gluons~\cite{zfp:c68:137,pr:d54:5523,pl:b478:101,hep-ph-0107068}. 
The exchange of the gluon ladder has also been 
computed~\cite{zfp:c68:137,pr:d53:3564,pl:b375:301,hep-ph-0107068}
in the leading logarithm approximation (LLA). 
The transition of a $q\qbar$ pair into a meson is, however, a 
non-perturbative
phenomenon that must be parameterised by the meson wave-function 
derived from lattice calculations and sum rules.

In the perturbative regime,
the cross section for the process $\gamma p \to V Y$ is given by
\begin{equation}
\frac{\diff^2 \sigma_{\gamma p \to V Y}}{\diff t \diff x} = 
g(x,-t)\frac{\diff \sigma_{\gamma g \to V g}}{\diff t}
 + \sum (q(x,-t)+\bar{q}(x,-t)) 
\frac{\diff \sigma_{\gamma q \to V q}}{\diff t}, 
\label{eq-my_pQCD}
\end{equation}
where $x$ is the fraction of the proton momentum carried by the 
struck 
parton and $g(x,-t)$, $q(x,-t)$ and $\bar q(x,-t)$ are the 
gluon, quark
and antiquark density functions, respectively.
The cross section for the production of a vector meson through 
the interaction 
with a gluon (\fig{feynman}(c)) is about five times larger 
than through
the interaction with a quark (\fig{feynman}(b)). 
At low $x$, therefore, the process $\gamma p \to V Y$ 
should be dominated by the reaction in which a gluon from 
the proton is struck.

In the two-gluon approximation,  the cross section
of Eq.~\eq{my_pQCD}
does not depend on $W$ for any given 
$t$, i.e. $\alpha^\prime = 0$.
In the LLA, this cross section has a power-law dependence on $W$,
although 
$\alpha^\prime$ is expected to be small, 
$\alpha^\prime \lesssim 0.1\gev^{-2}$ for 
$-t \gtrsim 1 \gev^2$~\cite{pl:b375:301},
decreasing with increasing $-t$ and reaching zero at 
asymptotically large $-t$ values.

A previous ZEUS publication~\citeHIGHT95 showed that 
$\diff \sigma/\diff t$ 
for the $\gamma p \to \rho^0 Y$ and $\gamma p \to \phi Y$ processes
were larger than those obtained by perturbative calculations 
performed in the two-gluon approximation~\cite{pr:d54:5523}.
These calculations 
have been further developed by 
Ivanov et al.~\cite{pl:b478:101}, who
proposed that a large
contribution would arise from $q\qbar$ fluctuations 
in a chiral-odd spin configuration. 
In this case, the quark mass appearing in the $\gamma V$ coupling 
is interpreted as a current quark mass, and therefore  
the quark mass in the light-meson wave-function is neglected. 
In such an approximation, 
$\diff \sigma/\diff t$ for $\rho^0$ and $\phi$ mesons
should have the same shape for $-t \gtrsim 1 \gev^2$. In addition, 
the cross sections should exhibit a power-law dependence on $t$,
$\diff \sigma/\diff t \propto (-t)^{-n}$; for $x>0.01$,  
the exponent $n\approx 3.8$ can be estimated from the 
Ivanov et al. calculation~\cite{pl:b478:101} for a fixed 
value of the strong coupling constant, $\als$, or 
$n\approx 4.8$ using the leading-order form for 
$\als(-t)$~\cite{epj:c5:461}.

In another two-gluon calculation, that of Forshaw and 
Ryskin~\cite{zfp:c68:137}, 
the quark mass appearing in the $\gamma V$ coupling is interpreted 
as the constituent quark mass. Such models predict a dip in 
$\diff \sigma/\diff t$ at $-t \approx M_V^2$, where $M_{V}$ is the 
meson mass. 

A LLA BFKL~\cite{jetp:45:199,*sovjnp:28:822} calculation has been
carried out by 
Bartels et al.~\cite{pl:b375:301}
for heavy mesons. The results
are expanded in terms of $\ln{(xW^2/W_0^2)}$, where 
$W_0^2=M_V^2-t$ is assumed.
This model predicts that 
$\diff \sigma/\diff t \propto (-t)^{-n}$, where $n$ is 
a function of the ratio $-t/M_{V}^2$.
For the $t$ range covered by the $J/\psi$ analysis of this paper, 
an average value of $n\approx 1.8$ is predicted for a 
fixed value of $\als$.
The previous ZEUS  measurement of the 
$\gamma p \to J/\psi Y$
cross section~\citeHIGHT95 shows that the prediction of this model, 
using $\als=0.2$,  agrees well with the
data. However, the uncertainties due to the choice 
of $\als$ and the $W_0$ scale are 
large.

In a recent paper, Forshaw and 
Poludniowski~\cite{hep-ph-0107068} have extended the BFKL
models of Forshaw and Ryskin~\cite{zfp:c68:137} and 
Bartels et al.~\cite{pl:b375:301}, and have given 
cross sections for the $\rho^0$, $\phi$ and $J/\psi$ mesons.
In this model, which uses a fixed value for $\als$,
a delta-function meson wave-function was used 
and the quark mass appearing in the $\gamma V$ coupling was 
assumed to be the constituent quark mass for all three mesons.  

\subsection{Helicity structure and cross-section ratios}
The hypothesis of $s$-channel helicity conservation 
(SCHC)~\cite{pl:b31:387} 
for vector-meson production 
states that the helicity of the final-state  meson will be 
equal to that of the initial photon. 
Studies of the elastic photoproduction of light vector mesons at 
low $-t$~\citeVM_HERA show 
that SCHC holds in these soft diffractive 
processes.

In VDM, the cross section for vector-meson production
is proportional to the product of the coupling of the photon 
to the vector meson and the meson-proton scattering cross section. 
The $\gamma \to V$ coupling depends on the meson leptonic width,
$\Gamma_{V \to e^+ e^-}$, which depends on the meson mass,
the quark assignment and  the wave-function.
The SU(4) prediction, which ignores the differences in the masses
and the wave-functions of the mesons, is that the ratio of the 
production cross sections for vector mesons will be
\begin{equation}
\label{eq-su4}
 \rho^0 : \omega : \phi : J/\psi = 1:1/9:2/9:8/9 .
\end{equation}

In pQCD, the helicity of the final-state meson and the cross-section
ratios are sensitive to the photon polarisation and 
the meson wave-function. 
In the trivial case of a meson wave-function 
given by a delta function, which is a good approximation
for heavy vector mesons,
the helicity of the final-state vector meson is equal to 
that of the initial photon, and thus SCHC will hold.
In this case, the cross-section for transverse 
photons is given by 
$\diff\sigma/ \diff t \propto M_V^3\Gamma_{V \to e^+ e^-}$
~\cite{zfp:c68:137,pl:b375:301,hep-ph-0107068}.
Based on the measured values of $M_V$ and 
$\Gamma_{V \to e^+ e^-}$~\cite{epj:c15:1}, the ratio of the 
photoproduction cross sections for various vector mesons at 
asymptotic values of $-t$ becomes 
\begin{equation}
\label{eq-tr_to_tr}
 \rho^0 : \omega : \phi : J/\psi = 1:(0.8\times 1/9):(2.1\times 2/9):(56\times 8/9), 
\end{equation}
which is significantly higher than
the SU(4) prediction of Eq.~\eq{su4} for the $\phi$ and 
$J/\psi$ meson. 

A wave-function more appropriate for the light 
mesons~\cite{pr:d54:5523,pl:b478:101}
leads to the production of mesons in the helicity-$0$
state by transverse photons and therefore to a violation of SCHC. 
The ratios of the production cross sections for mesons in the 
helicity-$\pm 1$ states by transverse photons
are equal to those of Eq.~\eq{tr_to_tr}.
The production of mesons in the helicity-$0$ state
by transverse photons, which is expected to be dominant at large $-t$, is given by
$\diff\sigma/ \diff t \propto M_V\Gamma_{V \to e^+ e^-}$
rather than $M^3_V\Gamma_{V \to e^+ e^-}$ mentioned earlier.
In this case, the ratios of the photoproduction
cross sections at large $-t$ should be
\begin{equation}
\label{eq-tr_to_lo}
 \rho^0 : \omega : \phi : J/\psi = 1:(0.8\times 1/9):(1.2\times 2/9):(3.5\times 8/9), 
\end{equation}
which lie between the values of Eqs.~\eq{su4} and~\eq{tr_to_tr}.
\section{Experimental set-up}
\label{sec-exp}
The measurements were performed with the ZEUS detector
at the HERA $ep$ collider. The data, corresponding to an integrated 
luminosity of $25.0 \pm 0.4 \pbi$,
were collected in 1996 and 1997, when HERA operated with a 
proton-beam energy of 
$820 \gev$ and a positron-beam energy of $27.5 \gev$.

\Zdetdesc

\Zctddesc\ZcoosysfnBeta

\Zcaldesc

A proton-remnant tagger (PRT1)\cite{zfp:c75:421} was used to tag events
in which the proton dissociates. 
It consists of two layers of scintillation
counters at $Z=5.15$~m.
The pseudorapidity range covered by the PRT1 is \mbox{$4.3<\eta<5.8$}.

The photoproduction tagger (PT)\citeHIGHT95 is a small
electromagnetic calorimeter located close to the beam-pipe 
at \mbox{$Z=-44$}~m.
It detects positrons of energy between 21 and $26 \gev$
scattered
under very small angles (less than a few mrads).
The positron measured in the PT is used to tag photoproduction
events with a photon-beam energy in the 1.5--6.5~GeV range.
 
The luminosity is determined from the rate of the Bethe-Heitler
bremsstrahlung process $e^+p \to e^+\gamma p$, where the high-energy 
photon is detected in a lead-scintillator calorimeter (LUMI) 
located at $Z=-107$~m in the HERA tunnel\cite{acpp:b32:2025}.  

\section{Kinematics and decay distributions}
\label{sec-kin}
\begin{figure}[p]
\vfill
\begin{center}
\epsfig{file=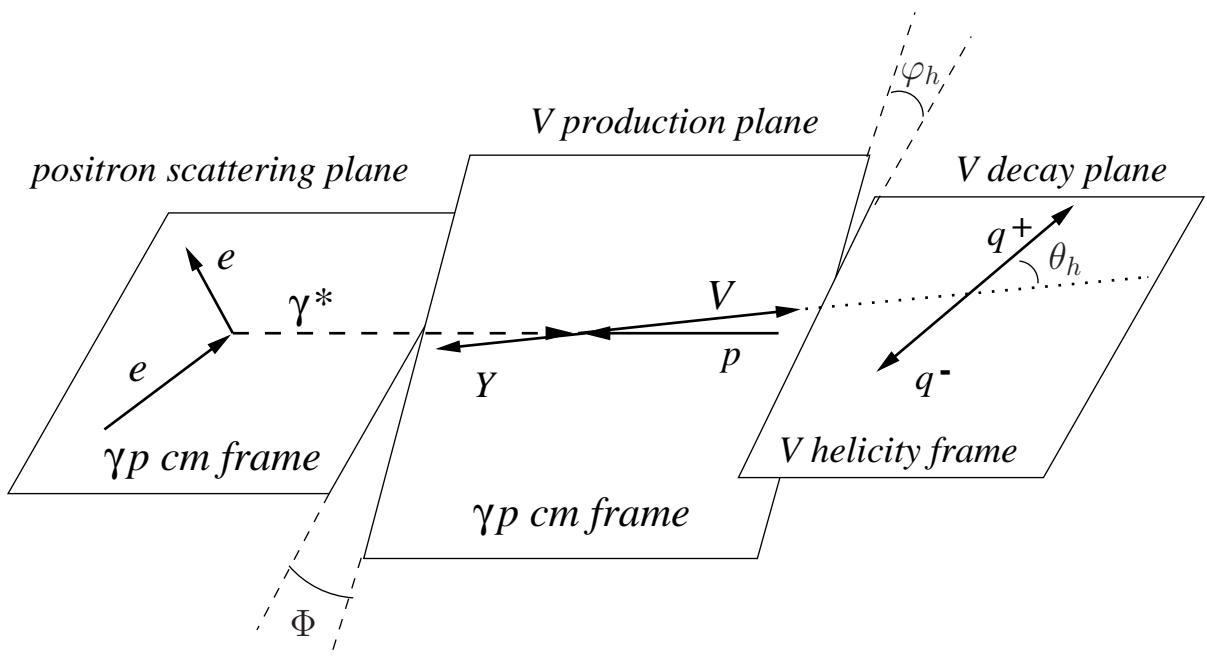,
width=1.00\textwidth}
\put(-347,11){\large $\Phi$}
\put(-116,222){\large $\varphi_h$}
\put(-60,150){\large $\theta_h$}
\end{center}
\caption{
Illustration of the angles used to analyse the helicity 
states of a vector meson 
(for a decay into two particles, $V \to \mbox{q}^+\mbox{q}^-$).}
  \label{fig-helicity}
\vfill
\end{figure}

\Fig{feynman}(d) shows a diagram of the reaction
$$e(k) p(P) \to e(k^\prime) V(v) Y(P^\prime),$$ 
where $V$ is a $\rho^0, \phi$, or $J/\psi$ meson and $k$, $k^\prime$, $P$,
$P^\prime$ and $v$ are the four-momenta of the incident positron, scattered
positron, incident proton, diffractively produced state $Y$ and vector meson,
respectively. The kinematic variables used to describe proton-dissociative
vector-meson production are:
\begin{itemize}
    \item $Q^2=-q^2 = -(k-k^\prime)^2$, the negative of the squared
     four-momentum of  the exchanged photon;
    \item $y=(P\cdot q)/(P\cdot k)$, the fraction of the positron energy 
     transferred to the photon in the rest frame of the 
     initial-state proton;
    \item $W^2 = (q+P)^2$, the squared
         centre-of-mass energy of the photon-proton system;
    \item $t=(P-P^\prime)^2 = (v-q)^2$, the squared four-momentum 
          exchanged at the hadronic vertex;
    \item $x$, the fraction 
          of the proton's momentum carried by the scattered parton. 
          Assuming that 
          the exchanged object in the photon-proton scattering 
          couples to a single massless parton in the proton
\begin{equation}
          \label{eq-xdef} 
          x=\frac{-(P^\prime-P)^2}{2P(P^\prime-P)} = -t/(M_Y^2-M_p^2-t),
          \end{equation}
          where $M_p$ is the mass of the proton and 
          $M_Y^2 = (P^\prime)^2$ is the 
          squared mass of the diffractively produced state, $Y$. 
\end{itemize}
The angles used are shown in~\fig{helicity} and are:
\begin{itemize}   
    \item $\Phi$, the angle, in the photon-proton centre-of-mass frame,
          between the positron-scattering plane
          and the $V$-production plane. The latter contains
          the momentum vectors of the virtual photon and the vector meson;
    \item $\theta_h$ and $\varphi_h$, the polar and azimuthal angles of
          the positively charged decay particle in the $V$ 
          helicity frame. The helicity frame is the 
          rest frame of the meson with the meson direction in the 
          photon-proton centre-of-mass frame taken as the quantisation 
          axis. The polar angle, $\theta_h$, 
          is defined as the angle between  
          the direction of the positively charged decay particle
          and the quantisation axis. 
          The azimuthal angle, $\varphi_h$, is the angle between 
          the decay plane and the $V$-production plane.
\end{itemize}

Only the momentum vectors of the meson decay products were
measured. 
The momentum of the scattered positron (and hence $Q^2$), the 
angle $\Phi$, and the mass $M_Y$ were not measured. 

In the tagged photoproduction events, $Q^2$ ranges from 
$Q^2_{\rm{min}}\approx 10^{-9}~\rm{GeV^2}$ up to 
$Q^2_{\rm{max}} \approx 0.02 \gev^2$, with a median $Q^2$ 
of approximately $7 \times 10^{-6}~{\rm GeV^2}$~\citeHIGHT95.
The value of $Q^2$ was neglected
in the reconstruction of the other kinematic variables.

The variables $W$ and $t$ can be expressed as~\citeHIGHT95
\begin{equation}
    W^2 \approx 2 E_p (E - p_Z)  \ \ \mbox{and} \ \ \nonumber
    t \approx - p_T^2,
\end{equation}
where $E_p$ is the incoming proton energy and $E$ is the energy, 
$p_Z$ is the longitudinal momentum and 
$p_T$ is the transverse momentum of the produced vector meson
in the laboratory frame. 

Since $\Phi$ is not measured, the three-dimensional angular 
distribution was averaged over this angle, so that 
the $\pm 1$ photon-helicity states are not distinguished.
The normalised two-dimensional angular distribution can then 
be written in terms of spin density matrix elements as
\begin{eqnarray}
\frac{1}{\sigma}\frac{\diff^2\sigma}{\diff \cos{\theta_h} \diff \varphi_h} 
& = & \frac{3}{4\pi} \left\{  \frac{1}{2}  (1 \mp r^{04}_{00}) \pm
\frac{1}{2}(3r^{04}_{00} - 1)\cos^{2}{\theta_h} \right. \nonumber \\ 
& & \left. \frac{}{} \mp \sqrt{2} 
\mbox{Re}[r^{04}_{10}]\sin{2\theta_h}\cos{\varphi_h}
      \mp r^{04}_{1-1}\sin^{2}{\theta_h}\cos{2\varphi_h} \right\}, 
\label{eq-hel_2d_dis}
\end{eqnarray}
where the upper (lower) signs are for the $V$ decay to spin-$0$ (spin-$1/2$)
particles. 
Integration over $\theta_h$ or $\varphi_h$ produces 
the one-dimensional distributions
\begin{equation}
\frac{\diff \sigma}{\diff \cos{\theta_h}} \propto   1 \mp r^{04}_{00} \pm  
(3r^{04}_{00}-1)\cos^{2}{\theta_h}  
\label{eq-the_hel}
\end{equation}
and
\begin{equation}
\frac{\diff \sigma}{\diff \varphi_h} \propto 1 \mp r^{04}_{1-1}\cos{2\varphi_h} .
\label{eq-phi_hel}
\end{equation}

Since $Q^2$ is very small for this study, the contribution from 
photons in the helicity-0 state is negligible. In this case,
the matrix elements are related to the helicity amplitudes,
$T_{\lambda_V \lambda_\gamma}$, as follows~\cite{np:b61:381}:
\begin{equation}
\label{eq-sdme}
r^{04}_{00} \approx \frac{T^2_{01}}{T^2_{01}+T^2_{11}+T^2_{-11}} ,
\ r^{04}_{10} \approx
\frac{1}{2}\frac{(T_{11}T^\star_{01})+(T_{-11}T^\star_{01})}{T^2_{01}+T^2_{11}+T^2_{-11}} ,
 \ r^{04}_{1-1} \approx
\frac{1}{2}\frac{(T_{11}T^\star_{-11})+(T_{-11}T^\star_{11})}{T^2_{01}+T^2_{11}+T^2_{-11}} ,
\end{equation}
where $T_{11}$ is the helicity non-flip amplitude, $T_{01}$ is  
the helicity single-flip amplitude and $T_{-11}$ is the helicity
double-flip amplitude. 
If SCHC holds, only the $T_{11}$
amplitude is non-zero and therefore $r^{04}_{00}$, $r^{04}_{10}$ and $r^{04}_{1-1}$
should all be zero.
%
%
\section{Event selection}
\label{sec-eve}
Events were selected online with a three-level trigger 
system~\cite{zeus:1993:bluebook}.
The selection of the reaction $e p \to e V Y$ at large values of $-t$
required:
\begin{itemize}
\item at the first trigger level:
 \begin{itemize}
 \item  a minimum energy deposit of $1 \gev$ in the PT;
 \item  at least one track candidate in the CTD.
 \end{itemize}
 In addition, the following cuts were applied to further reduce the
 backgrounds from random coincidences between some activity in the CTD
 and a detected positron in the PT:
 \begin{itemize}
 \item  an upper limit of $1 \gev$ on the energy 
 deposited in the LUMI photon detector.
 This requirement suppresses bremsstrahlung events and rejects
 hard QED radiation;
 \item at least  0.5 $\gev$ of energy deposit in the EMC section of the RCAL.
 This cut rejects proton-gas events. However, it also reduces the acceptance 
 for  $e p \to e V Y$ events and therefore was used only when necessary;
 \end{itemize}
\item at the second trigger level:
 \begin{itemize}
 \item CAL timing consistent with an $ep$ collision;
 \item not more than three tracks in the CTD;
 \end{itemize} 
\item at the third trigger level:
 \begin{itemize} 
 \item  exactly two tracks in the CTD pointing
 to a common vertex within $-60 < Z < 60$~cm;
 \item the transverse momentum of at least one track candidate 
 greater than $0.8\gev$. This cut efficiently
 selects events with large values of $-t$.
 \end{itemize}
\end{itemize}

After the event reconstruction, the following requirements were imposed:
\begin{itemize}
\item an interaction vertex within $-40 < Z < 40$~cm 
      and a transverse distance within $0.7$~cm of the nominal interaction
      point (IP);
\item exactly two tracks with opposite charges, each associated
      with the reconstructed event vertex, and each with 
      pseudorapidity $|\eta|< 2.1$
      and transverse momentum greater than 150~$\mev$ for the $\rho^0$, 
      400~$\mev$ for the $\phi$, and 500 $\mev$
      for the $J/\psi$ candidates;
\item energy deposits in the CAL (excluding FCAL cells with $\eta> 2.1$),
      not associated with 
      tracks, smaller than $250\mev$. This cut rejects 
      events with an additional particle not associated 
      with either the vector meson or the proton-dissociative system.
\end{itemize}

In addition, $-t>1.1 \gev^2$ and $80<W<120 \gev$ were required to 
select a kinematic region of high acceptance. The final number 
of events, after all selection cuts, was 21414 
for $0.55<M_{\pi\pi}<1.3 \gev$, 2407 for $0.99<M_{KK}<1.06\gev$
and 214 for $2.7<M_{ll}<3.2\gev$. 
%
%
\section{Monte Carlo simulation and acceptance calculation}

\subsection{Monte Carlo simulation}
\label{sec-mcs}
The EPSOFT~\citeEPSOFTmk Monte Carlo (MC) generator was used
for the simulation of the reaction $\gamma p \to V Y$.
The particular version~\citeEPSOFTla used here had an
improved simulation of the final-state particle multiplicity.
The DIFFVM~\citeDIFFVM  MC generator 
was used for systematic checks. 
In these generators, $\gamma p$ interactions are simulated
assuming the exchange of a colourless object. The cross-section
$\diff^2 \sigma_{\gamma p \to V Y} / (\diff t \diff M_Y^2)$
was parameterised using an exponential
$t$ distribution and a $1/M_Y^2$ distribution; SCHC was also assumed. 
Because the MC samples thus generated did not give a sufficiently 
good description of
the data, they were reweighted as described in the next section.  
The only difference between the EPSOFT and DIFFVM generators
that is relevant for this analysis  is the fragmentation scheme 
of system $Y$:
\begin{itemize}      
\item in EPSOFT, the exchanged object is assumed to couple to the
whole proton (\fig{feynman}(a)), 
which subsequently fragments into a state $Y$. 
The particle multiplicities and the transverse momenta of 
the hadrons in $Y$ are 
simulated using parameterisations of $pp$
data~\cite{prl:48:1451,*rncim:6:1}, while the 
longitudinal momenta are generated with  
a uniform rapidity distribution;
\item in DIFFVM, the exchanged object couples to a single
quark (\fig{feynman}(b)) in the proton. The fraction 
of the proton's momentum carried by the struck quark is
given by Eq.~\eq{xdef}.
The struck quark and the diquark remnant are colour-connected 
and are fragmented into the final-state $Y$ by the
JETSET7.3\cite{cpc:82:74} program. 
\end{itemize}
As pointed out in~Section~\ref{sec-pqcd},
pQCD models predict that the process $\gamma p \to V Y$ is
dominated by the reaction involving a gluon (\fig{feynman}(c)) 
from the proton. This process was implemented 
as an option in DIFFVM. In the 
following sections, the DIFFVM sample where this option was selected
is called DIFFVMg, while the DIFFVM sample in which
the proton emits a quark is called DIFFVMq.

The response of the ZEUS detector to  generated particles was
simulated using a program based on 
GEANT3.13~\cite{tech:cern-dd-ee-84-1}. 
The generated events were processed through the same selection and
reconstruction procedures as the data.

\subsection{Modelling of the dissociative-system $Y$}
\begin{figure}[p]
\vfill
\begin{center}
\epsfig{file=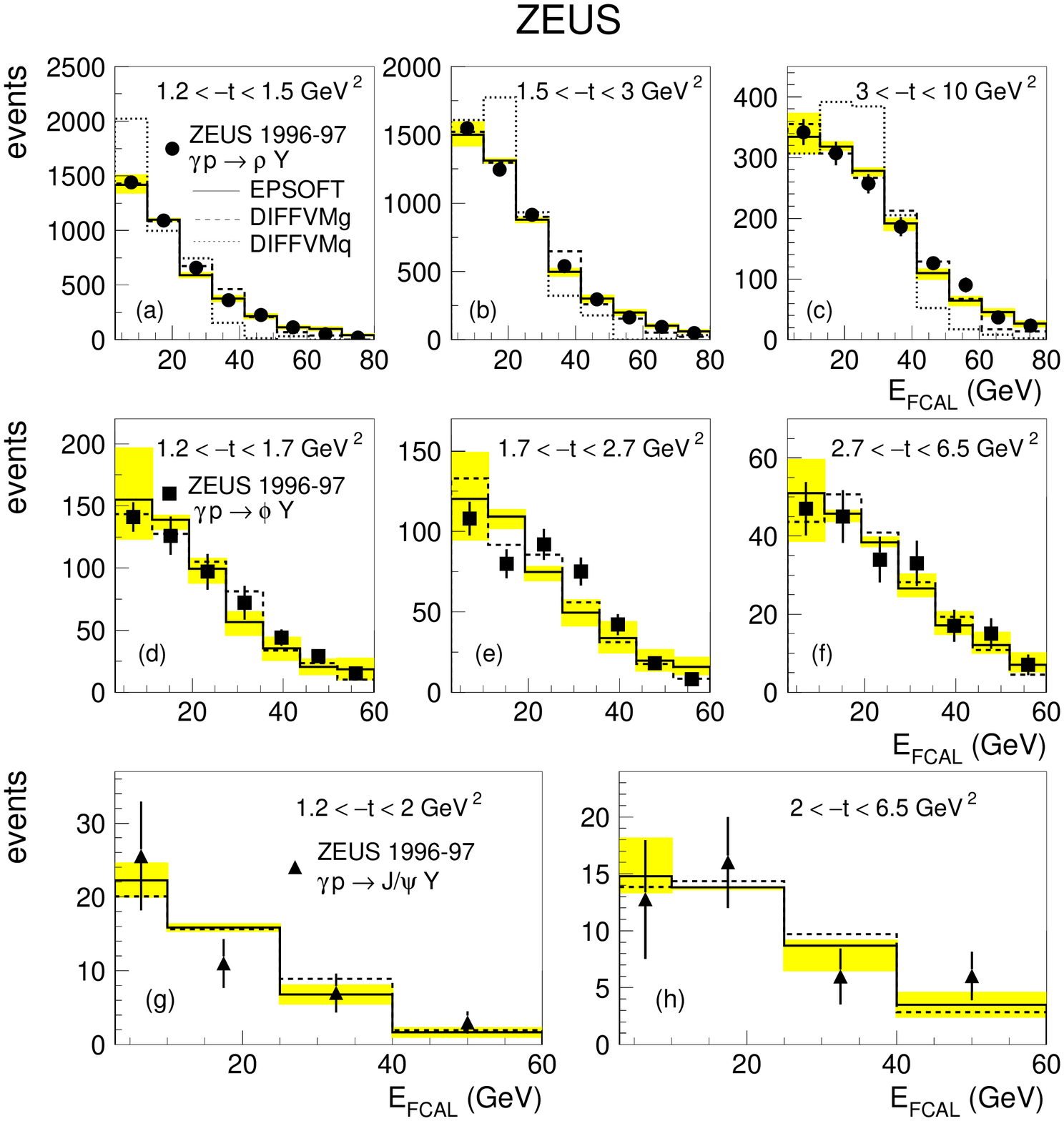,
width=1.00\textwidth}
\end{center}
  \caption{The distributions of the FCAL energy for: (a,b,c) $\rho^0$; 
 (d,e,f) $\phi$; (g,h) $J/\psi$ candidate samples in different $-t$ 
  regions.   The symbols 
  represent the data and the histograms indicate the results of the 
  simulation using EPSOFT (solid line), DIFFVMg (dashed lines)
  and DIFFVMq (dotted lines shown only for the $\rho^0$). The
  shaded bands represent the size of the correlated uncertainties due 
  to the  modelling of the proton dissociation in EPSOFT. 
}
  \label{fig-beta_comp}
\vfill
\end{figure}

The FCAL energy distributions are sensitive to
the fragmentation schemes of the system $Y$ 
as well as to
the shape of the generated $M_Y$ distribution.
The latter was reweighted with a function
$(1/M_Y^2)^{\beta(t) -1}$
to have the effective dependence
\begin{equation}
\frac{{\rm d}\sigma_{\gamma p \rightarrow VY}}{{\rm
d}M_Y^2} \propto \left( \frac{1}{M_Y^2} \right) ^{\beta(t)}.
\label{eq-triple}
\end{equation} 
The exponent $\beta(t)$ was chosen so as to reproduce 
the measured energy distributions in the FCAL
for the $\rho^0$ data sample: $\beta(t) = 0.93e^{0.11 t}$
for EPSOFT and $\beta(t) = 0.67e^{-0.1 t}$ for DIFFVMg.
These parameterisations are also valid for the $\phi$ and
$J/\psi$ mesons. The comparisons between the data and the MC simulations 
are displayed in~\fig{beta_comp} for several $t$ 
ranges and for the three mesons.
The agreement is satisfactory for DIFFVMg and EPSOFT but not
for DIFFVMq (shown only for the $\rho^0$ in~\fig{beta_comp}). 
This observation is consistent with
the pQCD expectation that the  photoproduction of 
vector mesons with proton dissociation at low $x$ is 
dominated by the reaction in which a gluon from the proton couples to
the vector meson. 
The DIFFVMg samples were used for systematic checks since 
the results do not change significantly 
when the mixture of DIFFVMq and DIFFVMg expected in Eq.~\eq{my_pQCD} was used.

\subsection{Acceptance corrections}
\label{sec-acc}
The overall acceptance is the product of the PT acceptance and that of the
main detector.
The geometric acceptance of the PT was evaluated with a program that
uses the HERA beam-transport matrices to track the positron through
the HERA beamline. The simulation was tuned so as to reproduce the tagging 
efficiency of the PT for Bethe-Heitler events, 
$e p \to e \gamma p$, triggered by a photon measured in the LUMI 
photon detector. This procedure was described previously~\citeHIGHT95. 
The variations of the position and tilt of the positron beam at the IP 
observed during data taking~\cite{thesis:klimek:2001} were also simulated.
 
\begin{figure}[p]
\vfill
\begin{center}
\epsfig{file=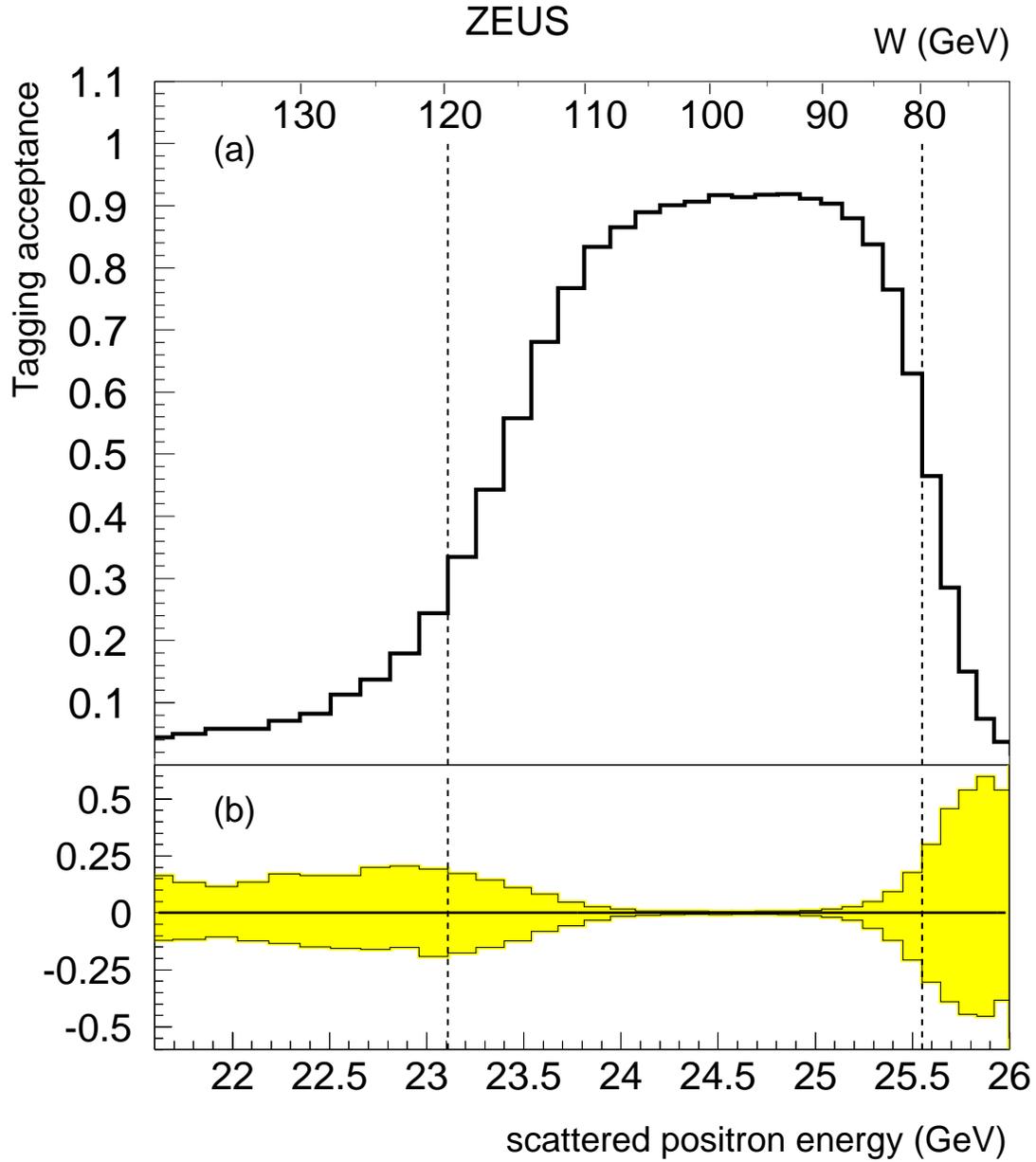,width=1.00\textwidth}
\vspace*{-6.5cm}
\end{center}
  \caption{
   The photoproduction tagging acceptance as a function of
   the energy of the scattered positron calculated
   using simulated events generated according to the equivalent-photon 
   approximation in the $Q^2<0.02 \gev^2$ range. 
   The shaded band represents the relative systematic 
   uncertainty of the tagging acceptance. 
   The dashed lines indicate the kinematic
   region used in this analysis. 
}
  \label{fig-tagger_acc}
\vfill
\end{figure}

The efficiency for photoproduction events was determined using the
geometric acceptance of the PT and events were generated according to
the equivalent-photon approximation~\cite{prep:15:181}. The
photoproduction tagging efficiency, calculated as a
function of the positron energy, $E_{e^\prime}$, is shown in
\fig{tagger_acc}. 
The range $80<W<120\gev$ 
was chosen so that the systematic uncertainty of the tagging 
efficiency, shown by the shaded area,    
does not dominate the total systematic uncertainty.
For the kinematic range used in this
analysis, the cross-section-weighted PT acceptance averages 
$80\pm 6$\%~\footnote{For these data, the PT was shifted 
closer to the positron beam, yielding a higher tagging efficiency 
than in the previous ZEUS measurement~\citeHIGHT95.}. 

\begin{figure}[p]
\vfill
\begin{center}
\epsfig{file=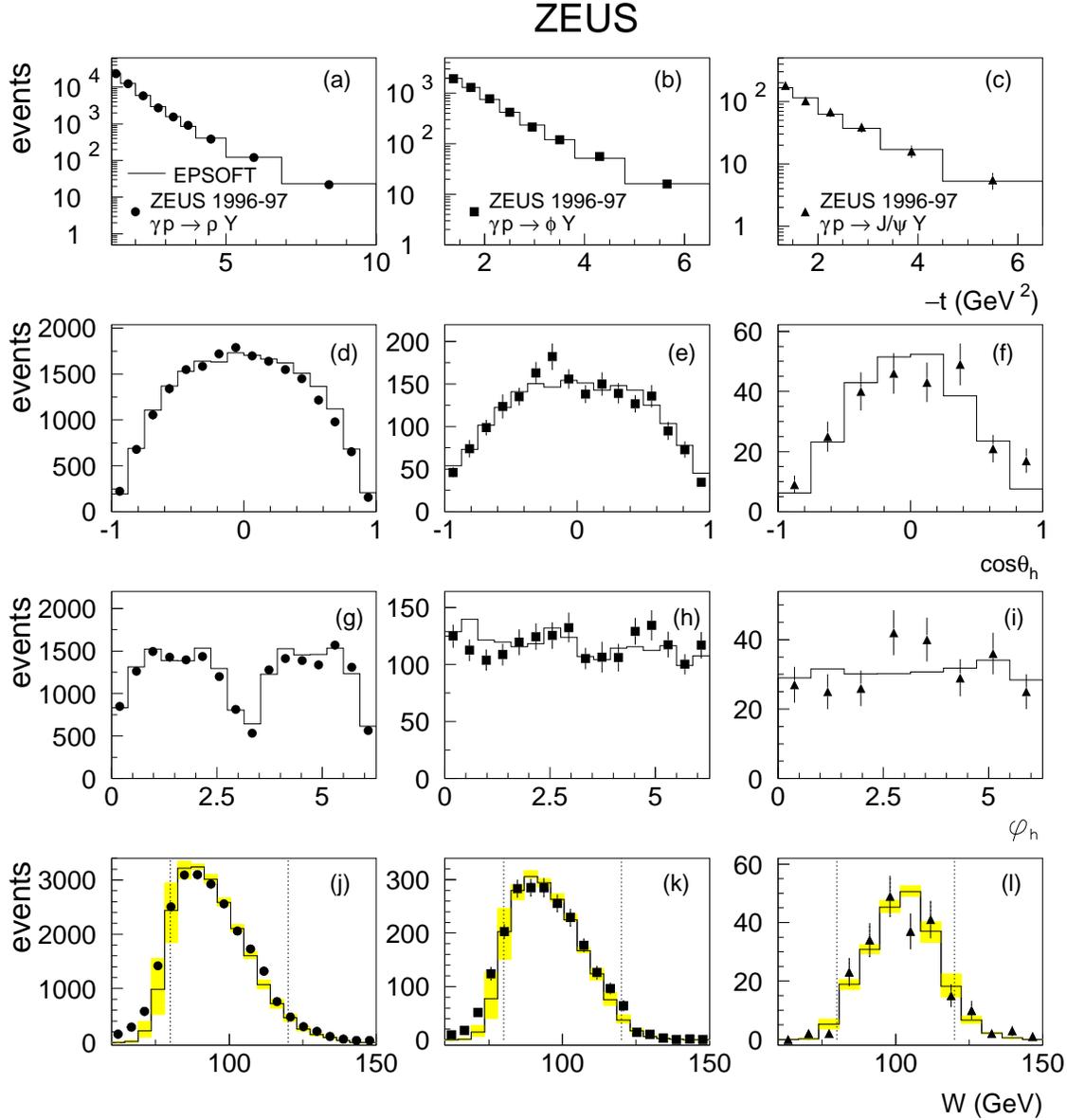,width=1.00\textwidth}
\end{center}
\caption{Comparison between the data and the EPSOFT MC (histograms) 
events for: (a,b,c) $-t$; (d,e,f) $\cos\theta_h$; (g,h,i) $\varphi_h$; 
(j,k,l) $W$, after all selection cuts (except the cut on $W$ in the case
of $W$ distributions). The MC is normalised to the data. 
The dashed lines (j,k,l) indicate the cuts on $W$ 
applied to select the final event sample. 
The systematic uncertainties due to the tagging acceptance 
are relevant in the case of $W$ distributions and are shown
as the shaded bands.
The three columns refer to the $\rho^0$, $\phi$, and $J/\psi$ samples, 
as indicated. 
} 
\label{fig-data_mc}
\vfill
\end{figure}

The decay-angle distributions at the generator level
were iteratively reweighted according to Eq.~\eq{hel_2d_dis} 
using the spin density matrix elements found in this study 
(see Section~\ref{sec-dec})
The spin density matrix elements for the $\rho^0$ and $\phi$ 
mesons were assumed to be linear functions of $t$;  SCHC was assumed
for the $J/\psi$ meson.
The generated $t$ distributions were reweighted with a function
$e^{P(t)}$, where $P(t)$ is a polynomial,
to reproduce those observed in the data. 
The resulting simulation agrees reasonably well with the data,
as shown in~\fig{data_mc}. 

Acceptance corrections evaluated on the basis of the reweighted 
EPSOFT sample, which gives the best description of the data,
were calculated separately for each bin of any given
variable.

On average, 6\% of $e p \to e V Y$ events were rejected by the trigger
due to accidental coincidence with bremsstrahlung photons in the LUMI
calorimeter. This effect was corrected for in the analysis.
%
%
\section{Backgrounds}
\label{sec-bac}
The dominant background sources, common to all channels,  came from
the double-dissociative, $\gamma p \to X Y$, and non-diffractive, 
$\gamma p \to$ anything, processes.  
Additional backgrounds contributing to individual channels are 
non-resonant $\pi^+\pi^-$ production for the $\rho^0$ 
sample, $\rho^0$~production for the 
$\phi$ sample, and inelastic Bethe-Heitler production of $e^+e^-$ and
$\mu^+\mu^-$ pairs for the $J/\psi$ sample.
These backgrounds were statistically subtracted using the fits to the 
invariant-mass distributions described in~\Sect{mas}.

The contribution of elastic vector-meson production 
($\gamma p \to Vp$) to the data was estimated from the fraction
of events, $F$, with a signal in one of the PRT1 counters 
above a threshold corresponding to the signal of a minimum-ionising particle. 
For elastic vector-meson production, such energy
deposits are absent. 
A comparison of the EPSOFT prediction
with the measured values of $F$  shows that the contribution
from the elastic process decreases from about $0.2$ at $-t=1.2 \gev^2$
to about $0.05$ at $-t=3.0 \gev^2$. The precision with which this
background is known is $0.01$ for the $\rho^0$, $0.03$ 
for the $\phi$ and $0.1$ for the $J/\psi$ meson.
This background was statistically subtracted bin by bin. 
For $-t > 3 \gev^2$, the contamination from the elastic reaction  
was consistent with zero and was neglected.

The contributions from the $\omega$ and $\phi$ mesons
(decaying to $\pi^+ \pi^- \pi^0$) in the $\rho^0$-candidate
sample were estimated using a MC simulation and the 
cross-section ratio given by the SU(4) prediction of $1/9$ 
for $\omega/\rho^0$ 
and the results of this analysis for $\phi/\rho^0$ 
(see Section~\ref{sec-rat}). This background was less than $3 \%$.
The contribution from the $\psi^\prime$ meson 
(decaying to $J/\psi+$ neutrals) in the $J/\psi$ candidate
sample was less than $3 \%$. These backgrounds contribute to
the overall normalisation uncertainty. 

\section{Systematic uncertainties}
\label{sec-sys}

The overall systematic uncertainties in the cross sections and density 
matrix elements in each bin were obtained by summing in quadrature
the uncertainties listed below:
\begin{itemize}
\item event selection:
   \begin{itemize}
   \item   the selection cuts, described in \Sect{eve}, were varied within the 
      resolutions of the cut variables;
   \item a different procedure to associate CTD tracks and CAL objects was used.
   \end{itemize}
      The effect on the cross sections was less than $\pm 10 \%$, 
      and less than $\pm 0.03$ in the spin density matrix elements;
\item signal extraction:
    \begin{itemize}
    \item the parameterisations of the signal shape were changed from the functional
      forms to the MC expectations;
    \item the ranges of the $M_{\pi\pi}$, $M_{KK}$ and $M_{ll}$ 
      mass distributions used in the fit procedure described in \Sect{mas}
      were varied by their resolutions.
    \end{itemize} 
      This resulted in changes of less than $\pm 10 \%$
      in the cross sections and less than $\pm 0.01$ in the spin 
      density matrix elements;
\item modelling of the dissociative-system $Y$:
     \begin{itemize}
        \item $\beta(t)$ in Eq.~\eq{triple}
              was changed to $\beta(t) = (0.93\pm0.07)e^{(0.11\pm0.06) t}$ for
              the $\rho^0$, $\beta(t) = (0.93\pm0.17)e^{(0.19\pm0.17) t}$ for
              the $\phi$ and $\beta(t) = 0.73\pm0.37$ for the $J/\psi$ meson.
              These ranges were chosen to maintain a satisfactory agreement 
              between the measured FCAL energy 
              distributions and the EPSOFT simulation;
        \item the fragmentation scheme of the system $Y$ was changed by
              using DIFFVMg instead of EPSOFT for the acceptance 
              corrections;
        \item $M_Y^2-M_p^2-t$ was used instead of $M_Y^2$~\cite{np:b77:240} 
              in Eq.~\eq{triple};
        \item the $1/M_Y^2$ dependence was reweighted to agree with 
              the pQCD prediction of Eq.~\eq{my_pQCD}~\footnote{The GRV98~\cite{epj:c5:461} 
              parton densities were used, but the results do not change when 
              other parameterisations are selected.}.
        \end{itemize} 
      The resulting uncertainty in the cross sections, coming mainly
      from the $\beta(t)$ variation, was less than $\pm 20 \%$. 
      All checks in this class change  the 
      generated $M_Y$ distribution. Therefore, to avoid double counting, 
      the maximum deviation from the nominal
      value was taken as the total systematic uncertainty of this class of
      systematic check.  
      Since the MC simulation assumes 
      factorisation of the proton and meson vertices, these systematic 
      checks result in no changes to the spin density matrix elements;
\item reweighting of the other MC distributions in EPSOFT:
    \begin{itemize}
    \item the $\theta_h$ and $\varphi_h$ distributions were reweighted 
      according to the SCHC
      predictions for the $\rho^0$ and $\phi$ mesons. For the $J/\psi$,
      the values $r^{04}_{00} = 0.13$ or $r^{04}_{1-1} = \pm 0.2$ were used,  
      corresponding to one standard deviation from 
      the SCHC values, as discussed in \Sect{dec}; 
     \item the $t$ and $W$ distributions were reweighted
      by a power-law function of $t$ and $W$, respectively,
      whilst maintaining satisfactory agreement between the data and the MC 
      distributions.
    \end{itemize}
      These modifications to the analysis procedure lead to
      changes in the cross section of less than $7 \%$ and less than $\pm 0.01$ 
      in the spin density matrix elements;        
\item calorimeter trigger efficiency:
    \begin{itemize}
    \item  the EMC RCAL trigger efficiency
      was varied within its uncertainty, as determined from
      subsamples of events triggered without the requirement of an
      energy deposit in the EMC section of the RCAL. 
    \end{itemize}
      This resulted in changes to the cross sections of less than $\pm 5 \%$
      and variations in the spin density matrix elements of up to $\pm 0.01$;
\item normalisation uncertainty:      
    \begin{itemize}  
     \item PT acceptance; the systematic uncertainty related to 
      the PT acceptance
      was evaluated by varying the inputs to the MC 
      simulation. The energy scale for photons
      measured in the LUMI detector was changed by 
      $\pm 50 \mev$~\cite{acpp:b32:2025}, 
      the horizontal position of the 
      PT by $\pm 0.3$mm, and the average position of the IP
      by $\pm 0.1$mm. These variations correspond  
      to the systematic uncertainties, as calculated from the 
      Bethe-Heitler event sample, discussed in \Sect{acc}. The effect on the cross sections 
      was $\pm 6 \%$;
     \item tracking-trigger efficiency; the uncertainty on the 
           CTD trigger efficiency was estimated from an
           independently triggered data sample and corresponds to an 
           uncertainty on the cross section of $\pm 4 \%$;
     \item PRT1 acceptance; the systematic uncertainty related to the 
           PRT1 acceptance was estimated from the difference between the
           fraction of events with a signal in the PRT1 in the data
           and in the EPSOFT simulation in the region of large $-t$,
           where the elastic contribution is expected to be negligible.
           This resulted in a systematic uncertainty of $\pm 3 \%$ 
           in the cross sections. This value was also used 
           in the low $-t$ region;
     \item other uncertainties;
           backgrounds from the $\omega$ and $\phi$ mesons in the 
           $\rho^0$ sample and from the $\psi^\prime$ meson
           in the  $J/\psi$ sample ($-3 \%$), as discussed in \Sect{bac} ;
           QED radiation ($\pm 2 \%$)~\citeHIGHT95;
           measurement of the integrated 
           luminosity ($\pm 1.8 \%$)~\cite{acpp:b32:2025};
    \end{itemize}
     The overall normalisation uncertainty of $\pm 10 \%$ was the sum in quadrature 
      of the above uncertainties. This uncertainty does not
      affect the spin density matrix elements.
\end{itemize}

\section{Results}
\label{sec-res}

Although $M_Y$ (and hence $x$) was not measured, the trigger and 
offline selection 
restricted $M_Y$ to be $\lesssim 25 \gev$, which
corresponds to $0.005 \lesssim x < 1 $ at $-t=2 \gev^2$ and 
$0.03 \lesssim x < 1$ at $-t=10 \gev^2$.  
To facilitate the comparison of the data with the pQCD predictions, 
the cross sections presented in the following sections were
evaluated in the fixed $x$ range $0.01< x< 1$, 
which approximately corresponds
to the $x$ range covered by the data at the average $t$ 
of this analysis.

\subsection{Cross-section determination}
\label{sec-csd}
The differential cross-section $\diff\sigma_{\gamma p \to V Y}/\diff \xi$ 
for the photoproduction
process $\gamma p \to V Y$ was obtained from the
cross section measured for the reaction $ep \to e V Y$ in each bin of 
the variable $\xi$($=t,\varphi,\cos\theta_h$) as
\begin{equation}
\frac{\diff \sigma_{\gamma p \to V Y}}{\diff \xi} =
\frac{\diff \sigma_{e p \to e V Y}}{\diff \xi} \frac{1}{\Phi_\gamma}
=\frac{N_V} {\Delta \xi \cdot {\cal L}\cdot \Phi_\gamma  \cdot B_V}, \nonumber
\label{gp_cros}
\end{equation}
where $\Phi_\gamma$ is the effective 
photon flux~\cite{prep:15:181}, $N_V$ is the background-corrected and acceptance-corrected 
number of 
vector mesons in the range 
$\Delta \xi$, extracted using the fits described in~\Sect{mas}, ${\cal L}$ is
the integrated luminosity and $B_V$ is 
the branching ratio of the vector-meson decay channel. 
The cross sections for  $\rho^0$ production
were integrated over the range 
$2M_\pi < M_{\pi\pi} < M_\rho + 5\Gamma_\rho$.

\subsection{Mass distributions}
\label{sec-mas}
\begin{figure}[p]
\vfill
\begin{center}
\epsfig{file=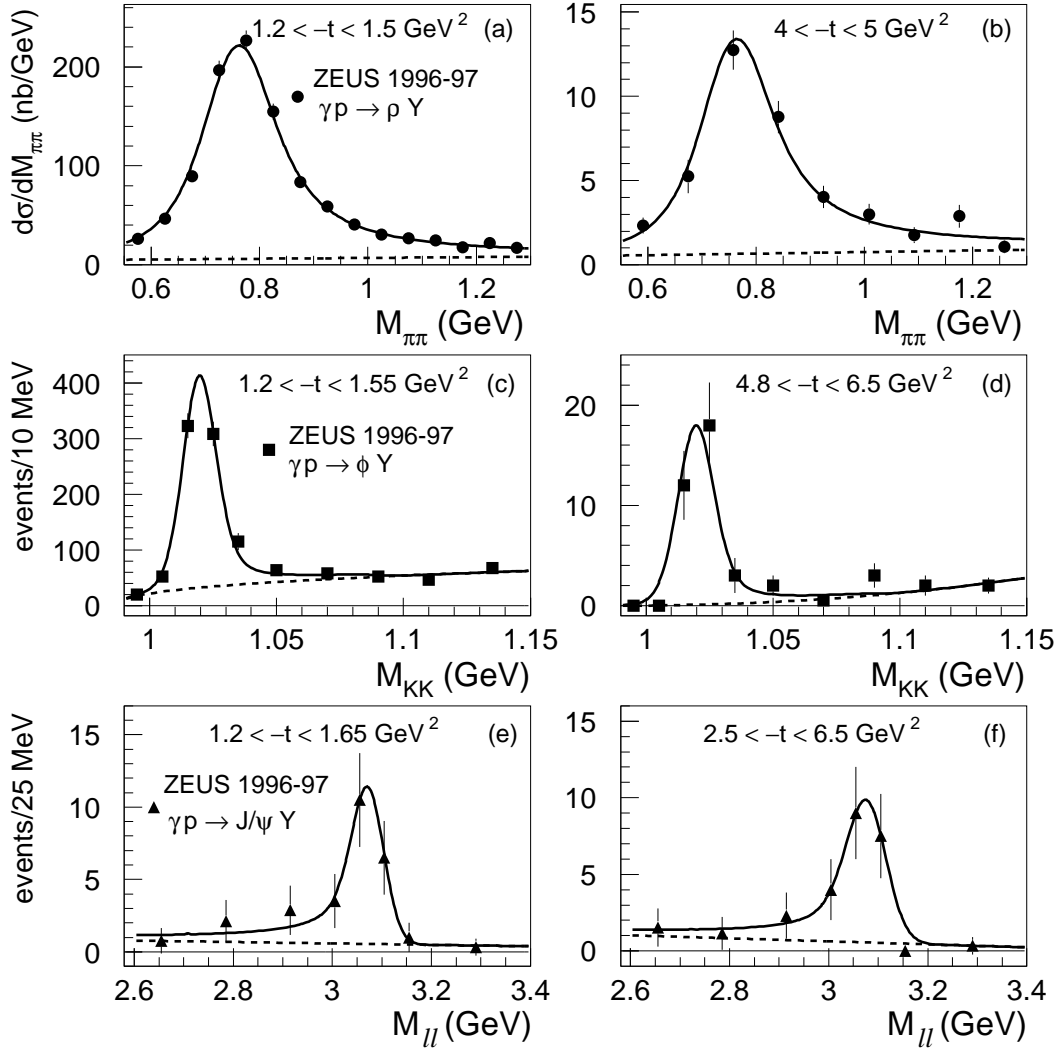,
width=1.00\textwidth}
\end{center}
\caption{(a,b) The differential cross-section
$\diff \sigma/\diff M_{\pi\pi}$ for the $\rho^0$ sample.
(c,d) The $M_{KK}$ mass distributions for the $\phi$ sample.
(e,f) The $M_{ll}$ mass distributions for the $J/\psi$ sample.
The symbols are the data for representative $t$ ranges and the 
solid curves indicate 
the result of the fits discussed in the text. The dashed curves 
show the background contributions. 
}
\label{fig-vm_mass}
\vfill
\end{figure}

The $\pi\pi$, $KK$, and $ll$ mass spectra for two representative 
$t$ ranges, together with the results of the fits discussed below, are 
shown in \fig{vm_mass}.

The acceptance-corrected $\pi\pi$ 
mass spectra for the $\rho^0$-candidate sample 
were fitted in the range 
$0.55 < M_{\pi\pi} < 1.3 \gev$ 
using the S\"oding parameterisation~\cite{pl:19:702}:
\begin{equation}
\frac{\diff\sigma}{\diff M_{\pi\pi}} = A_{\rho^0}^2
\left | \frac{ \sqrt{M_{\pi\pi} M_{\rho^0} \Gamma_{\rho^0}}}{M_{\pi\pi}^2-
M_{\rho^0}^2 +iM_{\rho^0} \Gamma_{\rho^0}}+ B/A_{\rho^0} \right |^2 + 
f_{PS}(M_{\pi\pi}) ,
\label{eq-soding}
\end{equation}
where $B$ is the non-resonant amplitude (taken to be
constant and real), $A_{\rho^0}$ is the normalisation
of the resonant amplitude, $M_{\rho^0}$ is the $\rho^0$ mass 
and $\Gamma_{\rho^0}$ is the momentum-dependent 
$\rho^0$ width~\cite{ncim:34:1644}.
The additional term  
$f_{PS} = A_{PS}(1+1.5M_{\pi\pi})$~\citeHIGHT95 was used 
to account for residual background from  the non-diffractive 
$\gamma p \to$ anything and the diffractive $\gamma p \to X Y$ processes. 
The $\chi^2/ndf$ for all the fits is satisfactory.
The ratio $|B/A_{\rho^0}|$ decreases from $0.13\pm0.02 \gev^{-1/2}$ 
at $-t=1.3 \gev^2$ to $0.05 \pm 0.03 \gev^{-1/2}$ at $-t=2.3 \gev^2$
and is consistent with zero above $-t=2.5 \gev^2$. 
The background contribution 
in the range $0.6 < M_{\pi\pi} < 1.3 \gev$, given by the 
integral of the function $f_{PS}$, increases from about 
$8\pm 2\%$ at $-t=1.3 \gev^2$ to $17 \pm 4\%$ at $-t=2.7 \gev^2$
and is consistent with 15\% above $-t=3.2 \gev^2$. 

The detector efficiency does not vary over the region
of the $\phi$ and $J/\psi$ peaks.
The $\phi$ and $J/\psi$ signals were, therefore, extracted 
from the uncorrected mass spectra.
The $KK$ mass spectra for the sample of $\phi$ candidates were 
fitted in the range 
$0.99 < M_{KK} < 1.15 \gev $
with the sum of a Breit-Wigner function convoluted with a Gaussian 
resolution function for the signal, and the function $(M_{KK}-2M_K)^\delta$ 
for the background. 
The $\chi^2/ndf$ for all the fits is satisfactory.
The background contribution in
the range $0.99 < M_{KK} < 1.06 \gev$ 
decreases from $25\pm 2 \%$ at $-t=1.3 \gev^2$ to 
$3\pm 2\%$ above $-t=4 \gev^2$. 

The $ll$ mass spectra for the $J/\psi$
candidate sample in the dilepton-invariant-mass range 
$2.6 < M_{ll} < 3.4 \gev $ were fitted with the sum of a Gaussian 
(for the muon decay
channel), a bremsstrahlung function convoluted with a Gaussian
(for the electron decay channel), and a background term linear in 
$M_{ll}$.
No muon or electron 
identification was performed;
the relative contributions of muon and electron
pairs were taken from the MC simulation.
The electron mass\footnote{The results do not change when the 
muon mass is used.} was used in calculating $M_{ll}$.
The $\chi^2/ndf$ for all the fits is satisfactory.
The background contribution in the range $2.7 < M_{ll} < 3.2 \gev$ 
is $25\pm 2\%$, approximately independent of the $t$ range under study.

Since the backgrounds depend on $t$, $W$, $\cos\theta_h$ 
and $\varphi_h$, each of the results presented in the following sections
was obtained by repeating the fits to the invariant
mass distributions in each kinematic bin.  
  
\subsection{Measurement of $\diff\sigma_{\gamma p \to V Y} / \diff t$}
\label{sec-dif}
\begin{figure}[p]
\vfill
\begin{center}
\epsfig{file=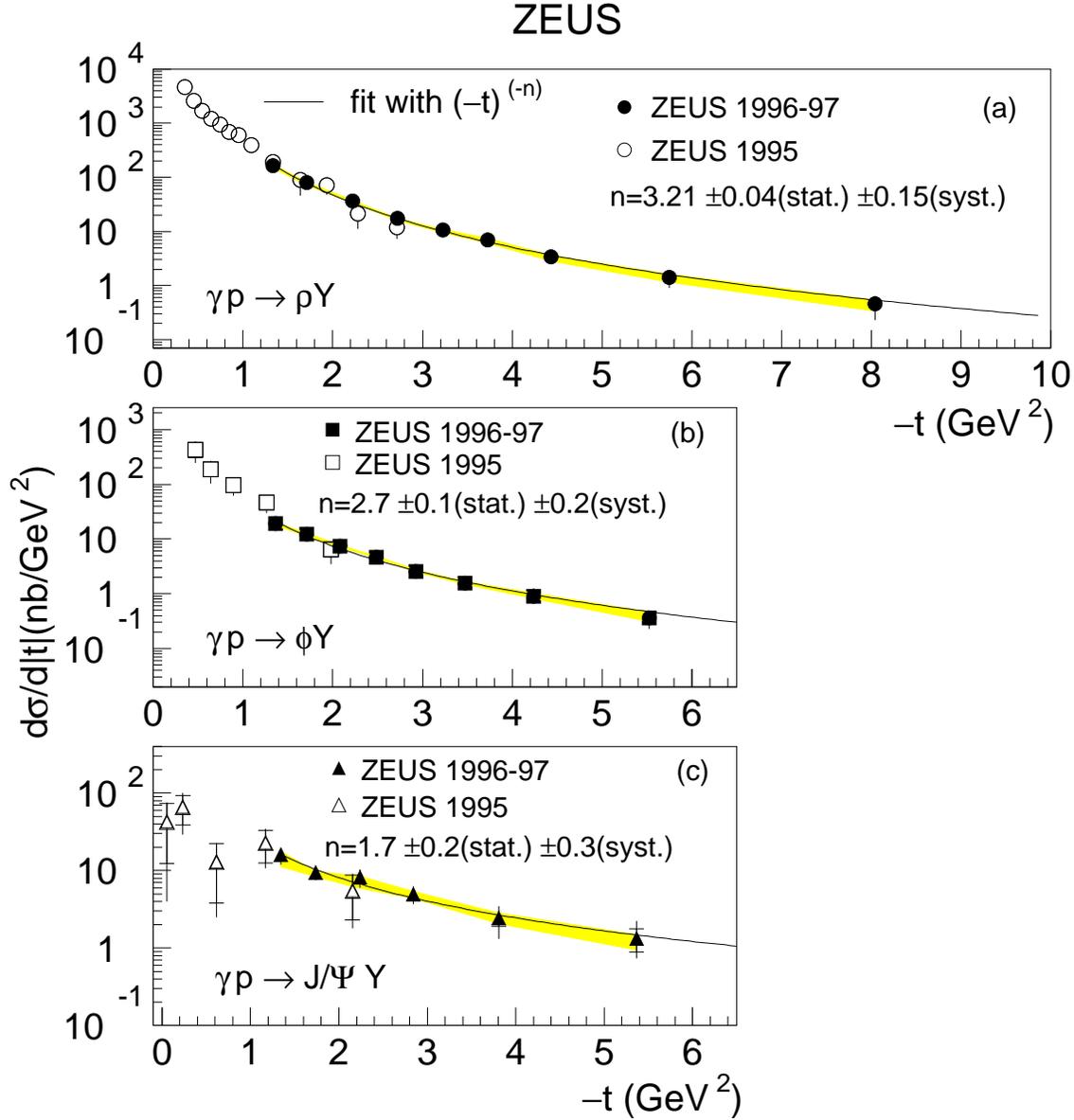,width=1.05\textwidth}
\end{center}
  \caption{The differential cross sections
  ${\rm d}\sigma_{\gamma p \to V Y}/{\rm d}|t|$ 
  in the range $80 < W < 120 \gev$ and $x>0.01$
  for: (a) $\rho^0$, (b) $\phi$ and (c) $J/\psi$.
  The inner bars indicate
  the statistical uncertainty and the outer bars represent the statistical and
  systematic uncertainties added in quadrature. The
  shaded bands represent  the correlated uncertainties due 
  to the  modelling of the hadronic-system $Y$.
  An additional correlated uncertainty of $\pm 10\%$ is not shown.   
  The open symbols correspond to the ZEUS 1995
  results~\protect\citeHIGHT95.
  The lines are the results of the fit to the data with the 
  function $A(-t)^{-n}$.}
  \label{fig-cross_rho_phi}
\vfill
\end{figure}

Tables~\ref{tab-rho_cros}~-~\ref{tab-psi_cros} and \fig{cross_rho_phi} show 
the differential cross sections for the $\rho^0$, $\phi$ and
$J/\psi$
mesons obtained from the present data 
together with results obtained in the ZEUS 1995~\citeHIGHT95
study of the lower $-t$ region.
In that analysis, the cross section
was determined for 
$M_Y^2/W^2<0.01$. In the kinematic region $-t\approx 1 \gev^2$ and 
$W\approx 100 \gev$, 
this limit is numerically equivalent to that used in the present 
study, so that the results may be directly compared.
The two measurements are consistent in the region of overlap.

The cross sections are well described 
by a power-law dependence, as expected for a hard
production mechanism~\cite{prl:63:1914,*pl:b284:123}. 
A fit to the present data with the function $A\cdot (-t)^{-n}$
gives 
$n = 3.21 \pm 0.04$(stat.)$\pm 0.15$(syst.) for the $\rho^0$, 
$n = 2.7 \pm 0.1$(stat.)$\pm 0.2$(syst.) for the $\phi$, and
$n = 1.7 \pm 0.2$(stat.)$\pm 0.3$(syst.)
for the $J/\psi$ meson. 

\begin{figure}[p]
\vfill
\begin{center}
\epsfig{file=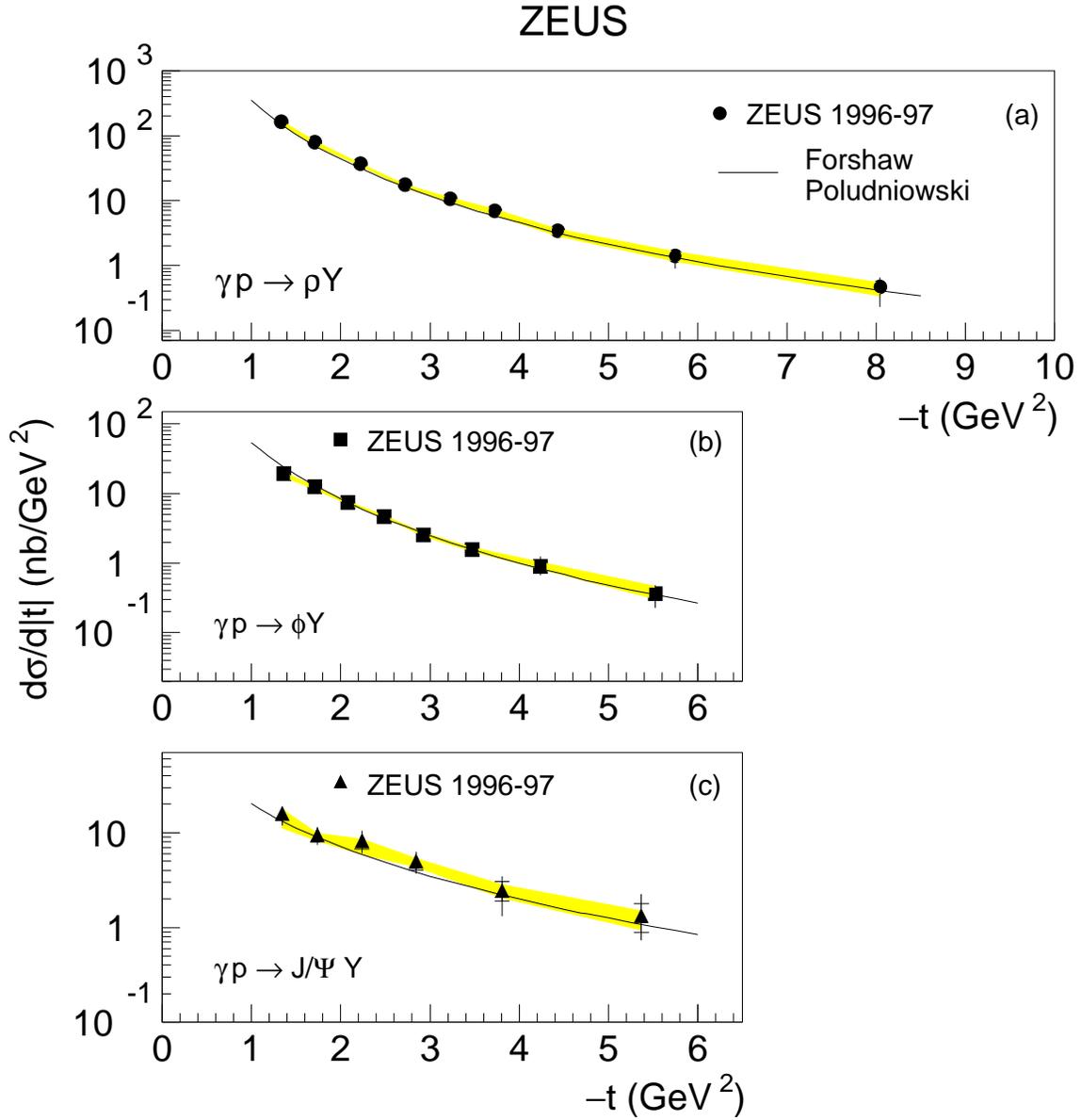,width=1.05\textwidth}
\end{center}
  \caption{The differential cross sections
  ${\rm d}\sigma_{\gamma p \to V Y}/{\rm d}|t|$  
  for: (a) $\rho^0$, (b) $\phi$, and (c) $J/\psi$ photoproduction 
  in the range $80 < W < 120 \gev$ and $x>0.01$.
  The solid curves show a
  pQCD calculation~\protect\cite{hep-ph-0107068}.
  The inner bars indicate
  the statistical uncertainty and the outer bars represent the statistical and
  systematic uncertainties added in quadrature. The
  shaded bands represent the size of the correlated uncertainties due 
  to the  modelling of the hadronic-system $Y.$
  An additional correlated uncertainty of $\pm 10\%$ is not shown.  
  }
  \label{fig-cross_psi}
\vfill
\end{figure}

The measured values of $n$ for the $\rho^0$ and $\phi$ mesons
are significantly smaller than the lower limit $n\approx 3.8$
predicted by Ivanov et al.~\cite{pl:b478:101}.
The value of $n$ for the $J/\psi$ meson agrees with
the Bartels et al.~\cite{pl:b375:301} BFKL 
prediction of 1.8.
The BFKL prediction of Forshaw and 
Poludniowski~\cite{hep-ph-0107068},
for the most natural choice of $\als=0.2$ and 
$W^2_0=M_V^2$, gives $\chi^2/ndf =0.5$ for all three mesons 
and is shown in \fig{cross_psi}.
These measurements establish that
perturbative QCD calculations are able to describe the large 
$-t$ regime. 
The dip at $-t\approx M_V^2$ predicted by the two-gluon 
exchange model of Forshaw and Ryskin~\cite{zfp:c68:137}
is not seen~\cite{hep-ph-0107068} in the data
for light mesons, indicating that this 
model is not applicable in the $-t\approx M_V^2$ region.  

Note that $n$ depends slightly on the $x$ range
used in the cross-section definition
both in the experiment and in the theory.
Hence the values 
quoted in this paper apply only to the cross section 
integrated over $x>0.01$. 

\subsection{Ratios of cross sections}
\label{sec-rat}
\begin{figure}[p]
\vspace*{0.5cm}
\begin{center}
\epsfig{file=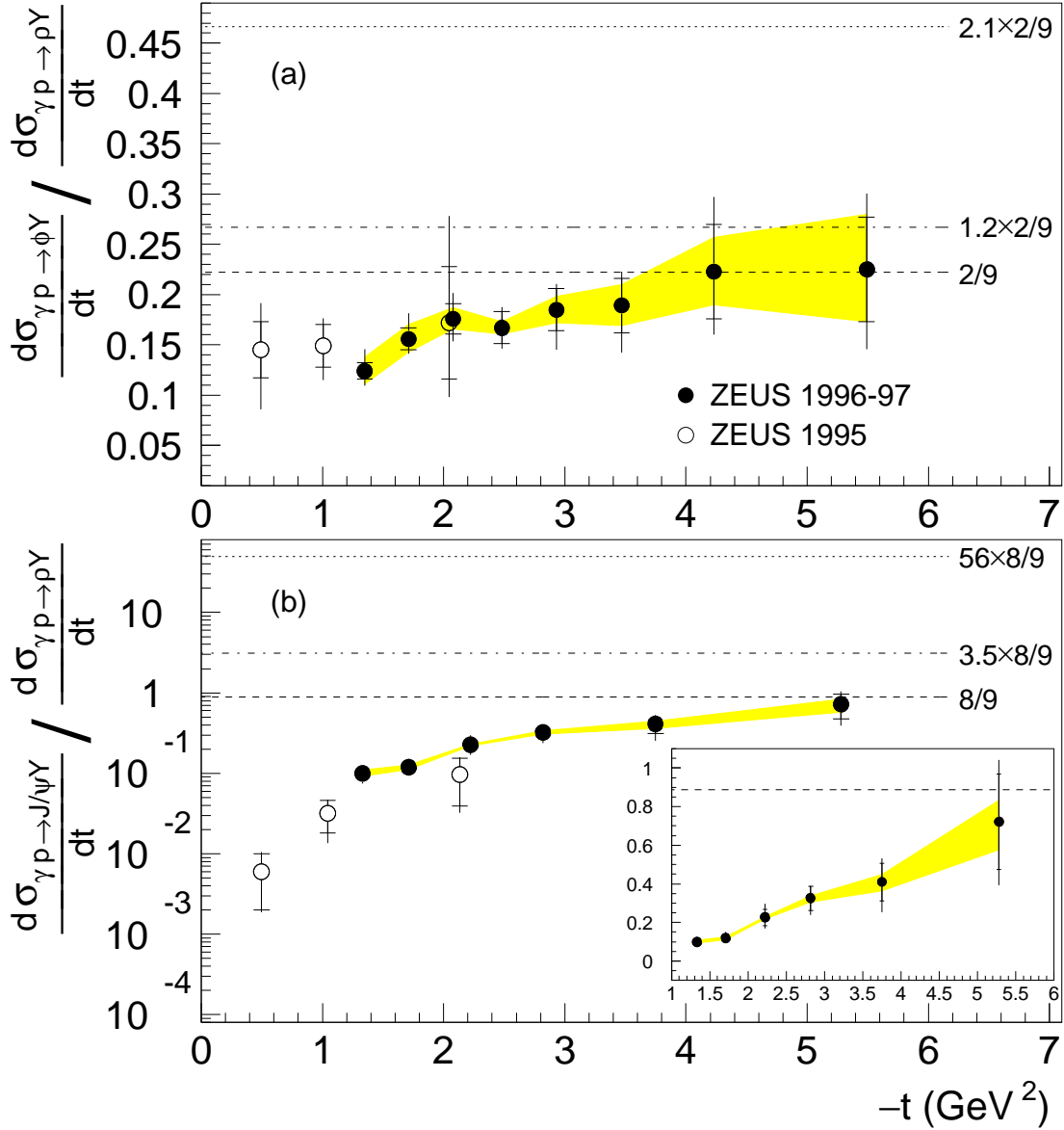,width=1.1\textwidth}
\end{center}
\vspace*{-0.5cm}  
  \caption{(a) The ratio of the cross sections ${\rm
  d}\sigma_{\gamma p \to V Y}/{\rm d}t$ for $\phi$ to $\rho^0$
  photoproduction. (b) The ratio of the cross sections ${\rm
  d}\sigma_{\gamma p \to V Y}/{\rm d}t$ for $J/\psi$ to $\rho^0$
  photoproduction. The inset shows this ratio on
  a linear scale. The inner bars indicate
  the statistical uncertainty, the outer bars represent the statistical and
  systematic uncertainties added in quadrature.  The
  shaded bands represent the size of the correlated uncertainties due 
  to the  modelling of the dissociative system, $Y$.
  The dashed lines correspond to the SU(4) predictions, while the 
  dotted and dashed-dotted correspond to the pQCD values given by 
  Eqs.~\eq{tr_to_tr} and~\eq{tr_to_lo}, respectively.
  The open circles are the ZEUS 
  1995 results~\protect\citeHIGHT95.}
  \label{fig-ratio}
\vfill
\end{figure}

The cross-section ratios
\begin{equation*}
r_{\phi}\equiv \frac{\diff\sigma_{\gamma p \to \phi Y}}
{\diff t} \mbox{\large /}
 \frac{\diff\sigma_{\gamma p \to \rho^0 Y}}{\diff t}
\end{equation*}
and 
\begin{equation*}
r_{J/\psi} \equiv \frac{\diff\sigma_{\gamma p \to J/\psi Y}}
{\diff t} \mbox{\large /}
\frac{\diff\sigma_{\gamma p \to \rho^0 Y}}
{\diff t}
\end{equation*}
are shown in Tables~\ref{tab-phi_rho} and \ref{tab-psi_rho}
and in \fig{ratio}, together with the values given in
Eqs.~\eq{su4},~\eq{tr_to_tr} and~\eq{tr_to_lo}.
A clear increase of $r_{\phi}$ and $r_{J/\psi}$
with increasing $-t$ is observed. 

The dependence
of $r_{\phi}$ and $r_{J/\psi}$ on $t$ suggests that the production 
mechanism  may become flavour independent at 
larger $-t$ values than covered in this analysis. 
In the framework of pQCD, this observation  
supports the hypothesis that the quark mass appearing in the 
$\gamma V$ coupling should be the constituent quark mass, since,  in 
this case,
the asymptotic value of $r_{\phi}$ should be reached at 
$-t\gg M_{\rho^0}^2, M_\phi^2 $. If the quark mass
is interpreted as the current quark mass, $r_{\phi}$
is expected to approach its asymptotic value at 
much smaller $-t$ values.

\begin{figure}[p]
\vfill
\vspace*{-0.5cm}
\begin{center}
\epsfig{file=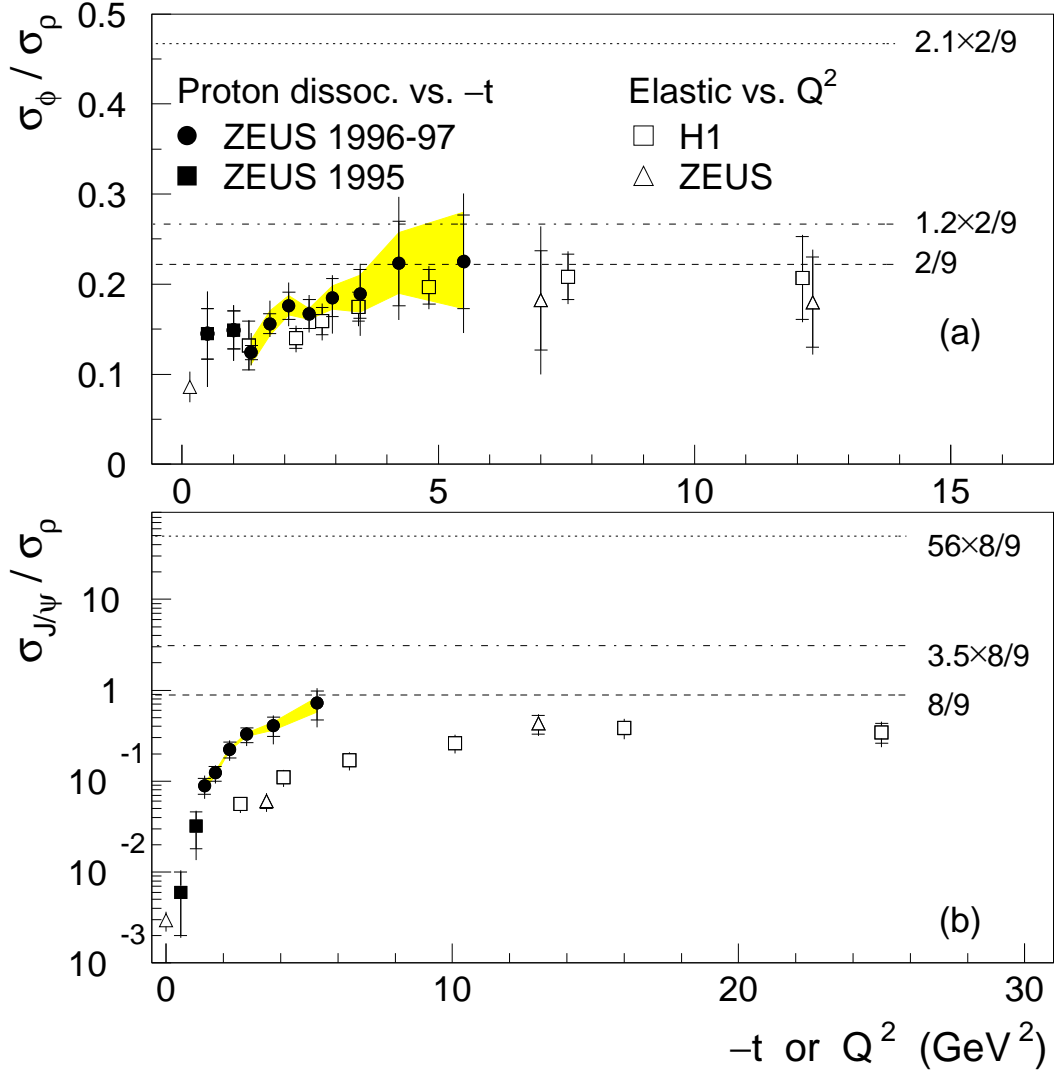,width=1.07\textwidth}
\end{center}  
\vspace*{-0.5cm}
  \caption{(a) The ratio of the $\phi$ to $\rho^0$ cross sections as
   a function of $-t$ or $Q^2$. The $\phi/\rho^0$ results as a function 
   of $-t$ for proton-dissociative photoproduction from this analysis 
   are shown with solid circles and those from the 
   ZEUS 1995~\protect\citeHIGHT95 measurement with the solid squares.
  The
  shaded bands represent the size of the correlated uncertainties due 
  to the  modelling of the dissociative system, $Y$.
   Open triangles at $Q^2\approx 0 \gev^2$~\protect\cite{pl:b377:259},
   $Q^2 =  7 \gev^2$~\protect\cite{pl:b487:273} and 
   $Q^2 =  12.3 \gev^2$~\protect\cite{pl:b380:220} represent the 
   $\phi/\rho^0$ ratio of the elastic cross sections as a function of 
   $Q^2$ from ZEUS, while the open squares represent those from 
   H1~\protect\cite{pl:b483:360}. 
(b) The ratio of the $J/\psi$ to $\rho^0$ cross sections as
   a function of $-t$ or $Q^2$. 
  The same convention for symbols as for $\phi/\rho^0$ ratio is used.
  Open triangles at $Q^2\approx 0 \gev^2$~\protect\cite{zfp:c75:215}
  and $Q^2 = 3.5, 13 \gev^2$~\protect\cite{epj:c6:603} represent the ZEUS
  measurements,
  while the open squares represent those of H1~\protect\cite{epj:c13:371,epj:c10:373}.     
  The dashed lines correspond to the SU(4) predictions, while the 
  dotted and dashed-dotted correspond to the pQCD values given by 
  Eqs.~\eq{tr_to_tr} and~\eq{tr_to_lo}, respectively.
}
  \label{fig-ratio_q2}
\vfill
\end{figure}

The cross-section ratios given in Eq.~\eq{tr_to_lo}, 
expected for the production of mesons in the helicity-$0$ state
by transverse photons at large $t$~\cite{pr:d54:5523,pl:b478:101},
are the same as those expected 
from pQCD for the elastic production of mesons in the helicity-$0$ state 
by longitudinal photons at  large $Q^2$~\cite{pr:d50:3134,shep:11:51}.
The cross-section ratios
for production of mesons in the helicity-$\pm 1$ states are expected 
to be larger than those
for production of mesons in the helicity-$0$ state.
\Fig{ratio_q2} compares the $t$ and $Q^2$ dependences of the
ratios for the proton-dissociative and for the elastic process, 
respectively;
$r_{\phi}$ lies systematically above $\sigma_{\gamma^* p \to \phi p}/ 
\sigma_{\gamma^* p \to \rho^0 p}$~\cite{pl:b483:360,pl:b487:273}  
but is compatible within the experimental uncertainties, while
$r_{J/\psi}$ for $-t> 3 \gev^2$
is about a factor of five larger than $\sigma_{\gamma^* p \to J/\psi p}/ 
\sigma_{\gamma^* p \to \rho^0 p}$
for $Q^2 > 3 \gev^2$ as measured by ZEUS~\cite{epj:c6:603} and
H1~\cite{epj:c13:371,epj:c10:373}~\footnote{The $J/\psi$ data from H1
have been scaled from $W=90\gev$ to $W=75\gev$ and the $\rho^0$ data
calculated at the same $Q^2$ values using a fit of the form 
$(Q^2+M^2_{\rho^0})^{-n}$~\cite{priv:meyer:2002}}. 
The helicity analysis, described in \Sect{dec}, shows that the production of mesons in the helicity-$\pm 1$ 
states dominates in the large $-t$ region. 
\subsection{The slope of the Pomeron trajectory}
\label{sec-wde}
The cross section given by Eq.~\eq{vdm}, integrated over
$M_Y$, can be expressed as
$$
\frac{\diff \sigma_{\gamma p \to V Y}}{\diff t} = F(t) W^{4\alpha(t)-4},
$$
where $F(t)$ is a function of $t$ only.
Assuming a linear form for the Pomeron
trajectory implies
\begin{equation}
\frac{\diff \sigma_{\gamma p \to V Y}}{\diff t} = F(t)  
W^{4\alpha(0)+4\alpha^\prime t-4}= F(t)  G(W)  W^{4\alpha^\prime t},
\label{eq-alphap}
\end{equation}   
where $G(W)$ is a function of $W$ only. 

The absolute $W$ dependence, and hence $\alpha(t)$, cannot be measured 
using the present data because of the large anti-correlated systematic
uncertainties of the PT acceptance as a function of $W$. 
However, these uncertainties do not affect the measurement of the 
relative changes of the $W$ dependence as a function of $t$,
which are sensitive to $\alpha^\prime$.
The slope of the Pomeron trajectory for the $\rho^0$ meson 
was obtained by fitting the form given by Eq.~\eq{alphap} 
to the measured cross sections using three $W$ times six $t$ bins.
A two-dimensional fit  gave  
$\alpha^\prime = -0.02 \pm 0.05 \mbox{(stat.)} _{-0.08}^{+0.04} 
\mbox{(syst.)} \gev^{-2}$ with $\chi^2/ndf =4.6/9$. 
A similar fit for the $\phi$ meson, performed in 
three $W$ times five $t$ bins,
gave $\alpha^\prime = -0.06 \pm 0.12 \mbox{(stat.)} 
_{-0.09}^{+0.05} \mbox{(syst.)} \gev^{-2}$ with $\chi^2/ndf =3.4/7$.
There were insufficient data to perform a similar analysis on the $J/\psi$
sample.

\begin{figure}[p]
\vfill
\begin{center}
\epsfig{file=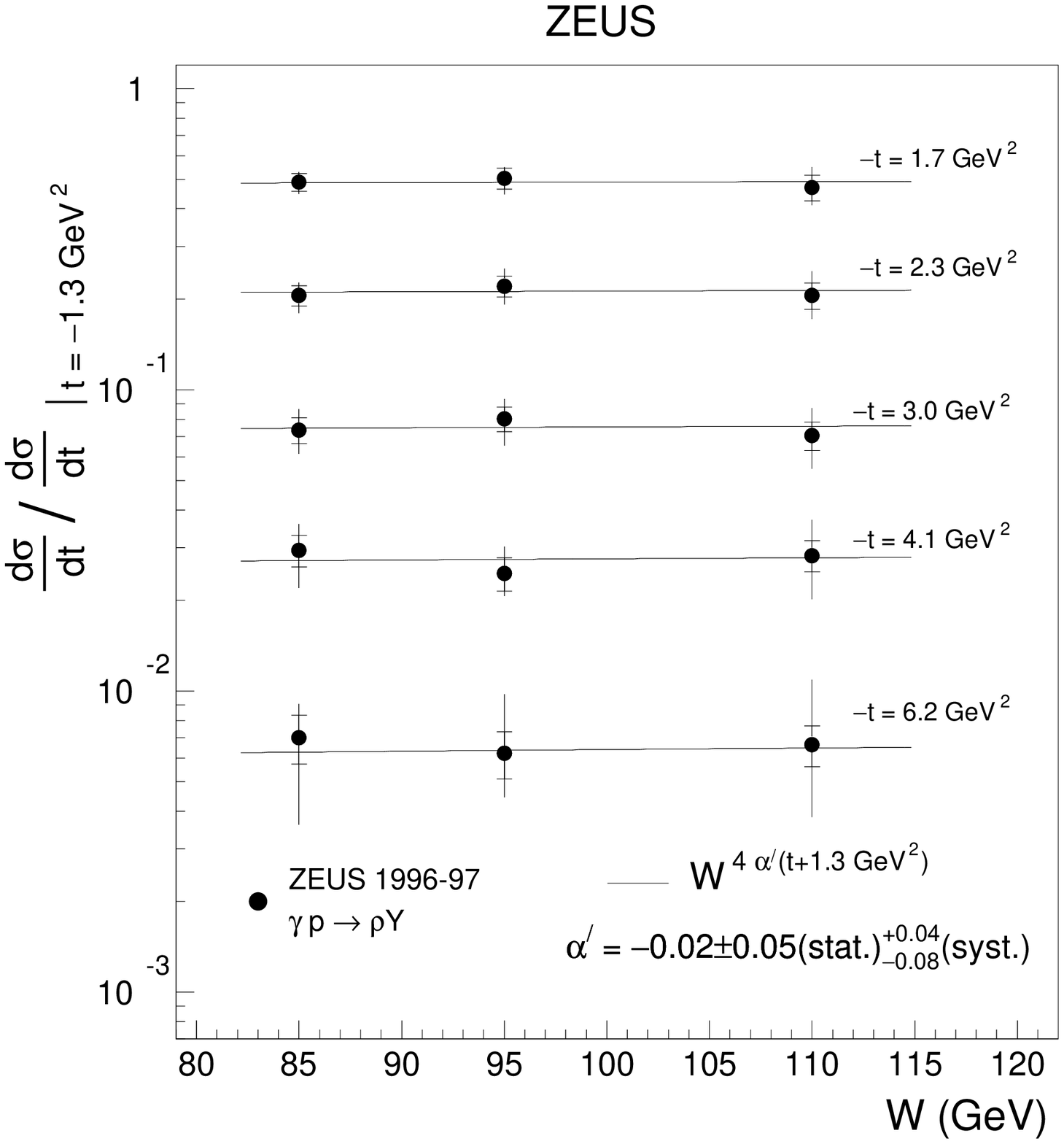,width=1.0\textwidth}
\end{center}  
  \caption{The ratios of $\rho^0$ production cross sections, 
$\frac{\diff \sigma}{\diff t} / \frac{\diff 
\sigma}{\diff t}|_{t=t_0}$,
for $-t_0=1.3$ {\rm GeV}$^2$, as a function of $W$ in five $t$ intervals.
The lines represent the result of the fit
with Eq.~\eq{alphap}.
  The inner bars indicate
  the statistical uncertainty and the outer bars represent 
  the statistical and
  systematic uncertainties added in quadrature.
}
  \label{fig-rho_delta}
\vfill
\end{figure}

The consistency of the $\alpha^\prime$ values with zero
implies that the $W$ dependence does not vary with $t$, and {\it vice versa}. 
This conclusion is manifest when looking at the $W$ dependence of the ratio of
differential cross sections 
$$
\frac{
\frac{\diff \sigma(W)}{\diff t}} 
{\frac{\diff \sigma(W)}{\diff t}|_{t=t_0}} 
\propto  
W^{4\alpha^\prime(t-t_0)},
$$
shown in \figand{rho_delta}{phi_delta}.
Values of $-t_0=1.3 \gev^2$ were chosen for the $\rho^0$
and $-t_0=1.5 \gev^2$ for the $\phi$ case.
 
\begin{figure}[p]
\vfill
\begin{center}
\epsfig{file=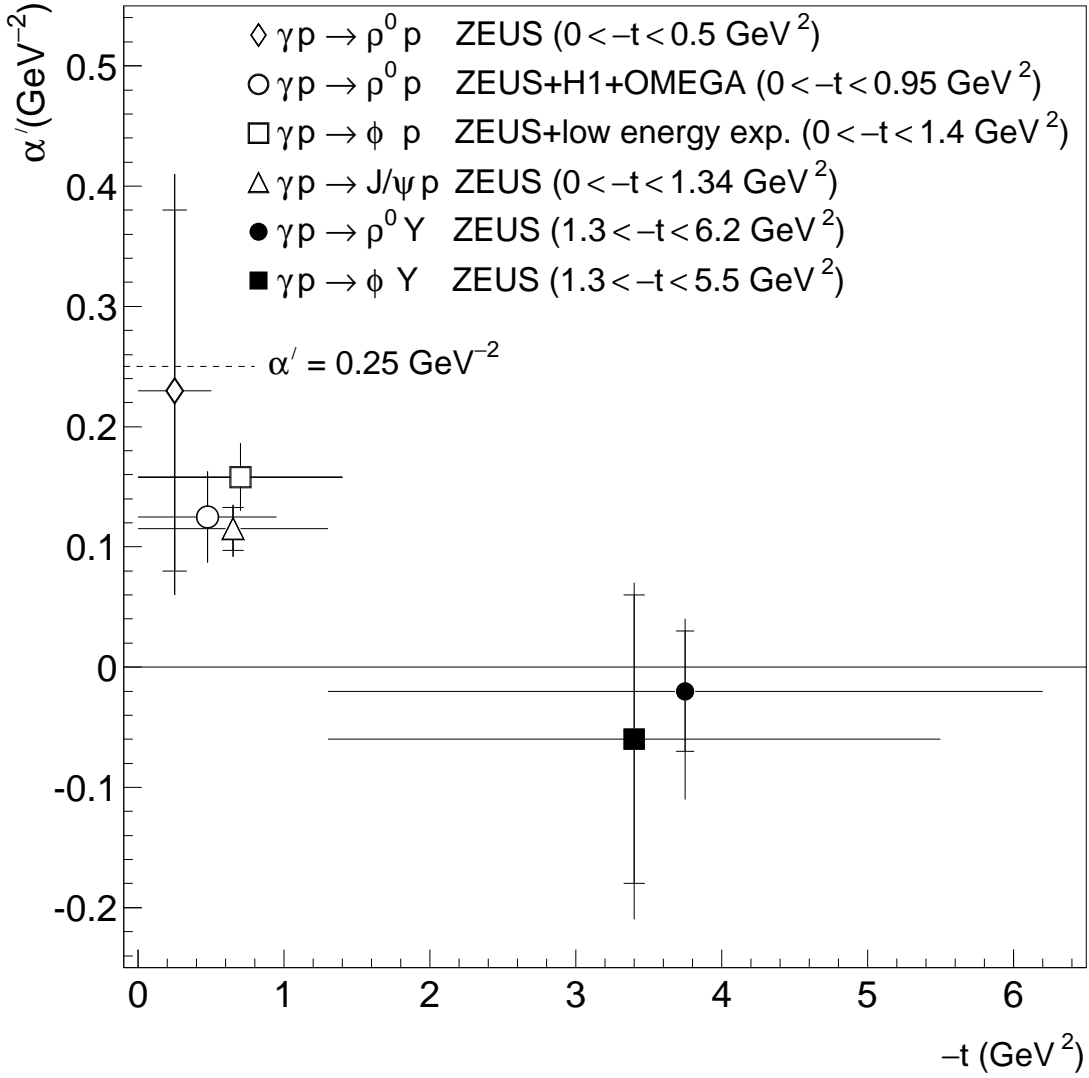,width=1.00\textwidth}
\end{center}  
  \caption{
Comparison of results on $\alpha^\prime$ for vector-meson production.
The result for the proton-dissociative photoproduction from this analysis 
are shown with solid symbols and those for elastic photoproduction
with the open symbols. The value of
$\alpha^\prime=0.25\gev^{-2}$ is
characteristic of soft hadronic processes.
  The horizontal bars correspond to the $-t$ range
   in which $\alpha^\prime$ is measured.
  The vertical inner bars indicate
  the statistical uncertainty and the outer bars represent 
  the statistical and
  systematic uncertainties added in quadrature.
}
  \label{fig-alpha-prime}
\vfill
\end{figure}

\Fig{alpha-prime} compares the values of $\alpha^\prime$ measured in this 
analysis with those obtained at lower $-t$.
For both $\rho^0$ and $\phi$ mesons, the values of $\alpha^\prime$ 
for $-t>1.3 \gev^2$ are smaller than the value 
$\alpha^\prime=0.25\gev^{-2}$ characteristic of soft hadronic processes
at $-t<0.5 \gev^2$ and also below the values 
$\alpha^\prime = 0.125 \pm 0.038\gev^{-2}$, 
$\alpha^\prime = 0.158 \pm 0.028\gev^{-2}$ and
$\alpha^\prime = 0.113 \pm 0.018\gev^{-2}$
obtained for $-t < 1.5\gev^2$,
from fits~\cite{epj:c14:213,hep-ex-0201043} to the elastic
photoproduction of $\rho^0$, $\phi$ and $J/\psi$ mesons, respectively.
This observation indicates that $\alpha^\prime$ decreases  
with increasing $-t$.
The small values of $\alpha^\prime$ obtained in this analysis
are consistent with the flatness of the Pomeron trajectory 
observed by UA8~\cite{np:b514:3} at large $-t$ and with
the expectation of the anomalous Regge exchange model by 
Kochelev et al.~\cite{npps:99a:24}. They are also consistent
with the expectation of $\alpha^\prime = 0$ characteristic of the 
two-gluon-exchange 
models~\cite{zfp:c68:137,pr:d54:5523,pl:b478:101,hep-ph-0107068}
and the expectation of $\alpha^\prime \lesssim 0.1\gev^{-2}$
of the BFKL model of Bartels et al.~\cite{pl:b375:301}.

\subsection{Decay angular distributions}
\label{sec-dec}

\begin{figure}[p]
\vfill
\begin{center}
\epsfig{file=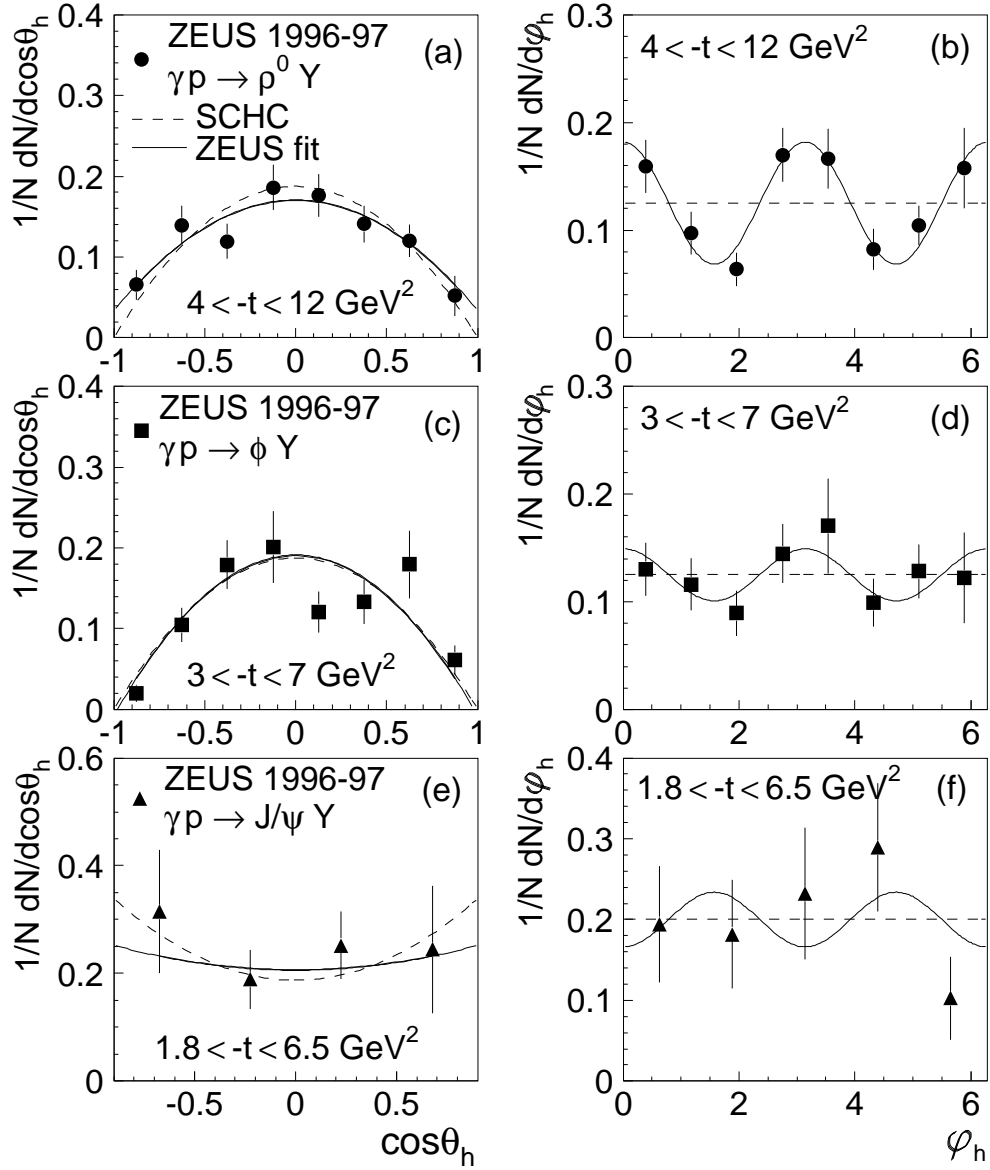,width=1.05\textwidth}
\end{center}  
  \caption{ The normalised background-subtracted and acceptance-corrected 
$\cos\theta_h$ and $\varphi_h$ distributions for proton-dissociative 
photoproduction of:
(a,b) $\rho^0$, (c,d) $\phi$ and (e,f) $J/\psi$. 
The symbols represent 
the data and the solid curves the result of the fits with 
Eqs.~\eq{the_hel} and~\eq{phi_hel}. 
The dashed curves are the SCHC predictions.
}
  \label{fig-vm_1dim_fits}
\vfill
\end{figure}

The $r^{04}_{00}$ and $r^{04}_{1-1}$ spin density matrix elements 
for the $\rho^0$, $\phi$, and $J/\psi$ were obtained by fitting Eqs.~\eq{the_hel} and~\eq{phi_hel} to the 
background-subtracted and acceptance-corrected data in several $t$ 
ranges. The $t$ ranges and the results of the fits are given in
Tables~\ref{tab-rhoSDM},~\ref{tab-phiSDM} and \ref{tab-jpsiSDM}.
As an example,~\fig{vm_1dim_fits} shows the $\cos\theta_h$
and $\varphi_h$ distributions with the results of the fits 
in the highest $-t$ bin used in the helicity analysis 
for each of the three mesons.
The flat $\varphi_h$ distribution shown by the dashed line 
disagrees ($\chi^2/ndf = 42/7$) with the $\rho^0$ data,
indicating a violation of SCHC.
The parameter $\mbox{Re}[r^{04}_{10}]$ for the $\rho^0$ and 
$\phi$ mesons was determined using a two-dimensional fit 
of Eq.~\eq{hel_2d_dis} in three and two $t$ intervals, respectively.
The results of the fits are  summarised in
\taband{rhoSDM}{phiSDM}.
The above results were also confirmed by a
decay-angle analysis~\cite{thesis:kowal:2002} 
performed in the transversity frame~\footnote{The spin-quantisation
axis in the transversity frame is chosen along the normal to the
meson production frame.}.
There were insufficient data to perform a two-dimensional fit
for the $J/\psi$ meson.

\begin{figure}[p]
\vfill
\begin{center}
\epsfig{file=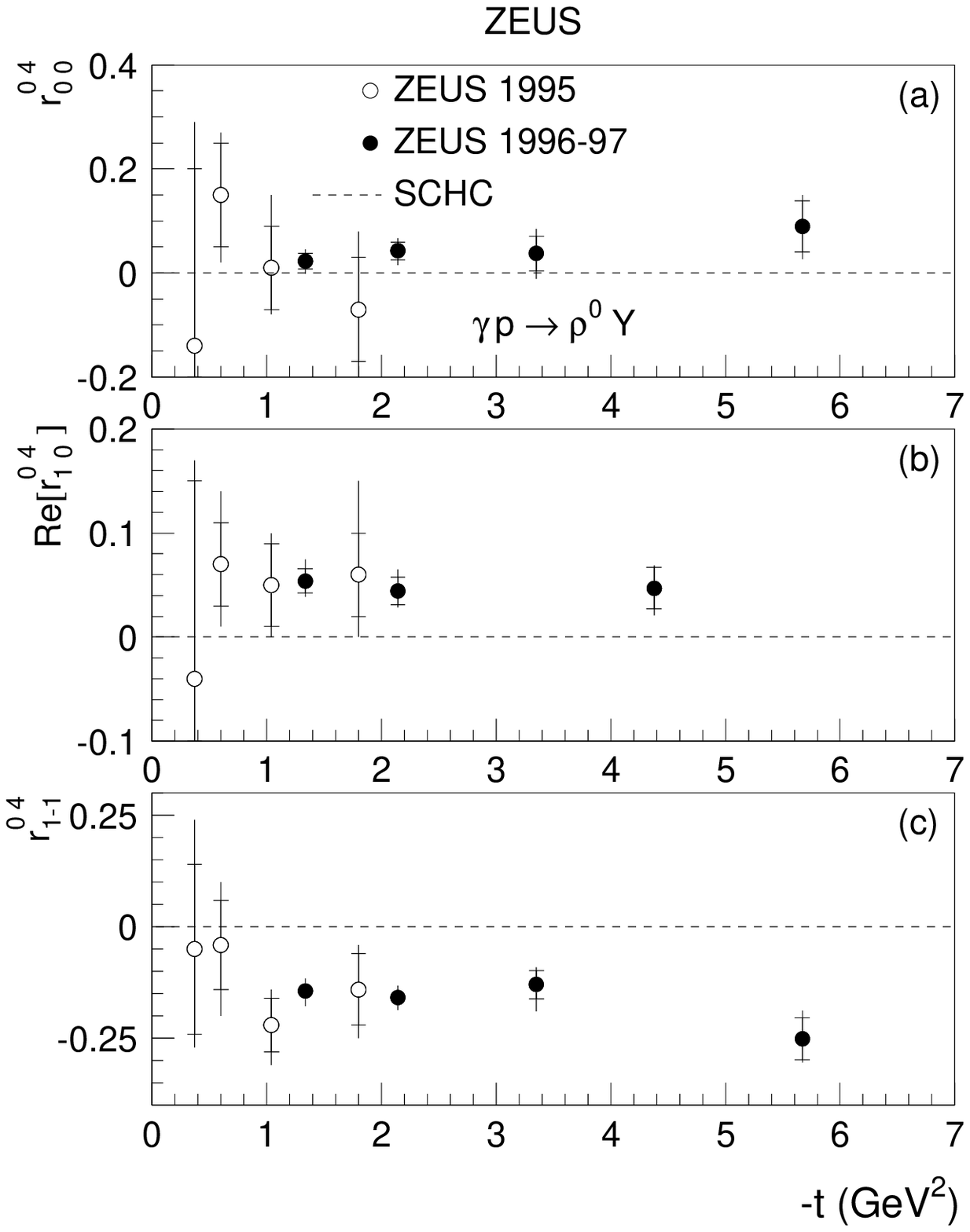,width=1.00\textwidth}
\end{center}  
  \caption{The fitted values of (a) $r^{04}_{00}$, (b) $\mbox{Re}[r^{04}_{10}]$
and (c) $r^{04}_{1-1}$ for proton-dissociative $\rho^0$ photoproduction 
as a function of $-t$.
  The inner  bars indicate
  the statistical uncertainty and the outer bars represent 
  the statistical and
  systematic uncertainties added in quadrature.
The open circles correspond 
to the ZEUS 1995 results~\protect\citeHIGHT95.
The SCHC prediction is shown as the dashed line.
}
\label{fig-rho_hel_res}
\vfill
\end{figure}

The fitted values of $r^{04}_{00}$, $\mbox{Re}[r^{04}_{10}]$,
and $r^{04}_{1-1}$ for the $\rho^0$ meson are displayed
in~\fig{rho_hel_res}.
The results are 
consistent with the ZEUS 1995~\citeHIGHT95 measurements in the
overlap region.
The small values  of $r^{04}_{00}$ indicate that the  
probability to produce $\rho^0$ mesons in the helicity-$0$ state 
from a photon in the helicity-$\pm1$ states 
is  $4 \pm 5\%$ at $-t=3.35 \gev^2$ and  $9 \pm 6\%$
at $-t=5.67 \gev^2$. However,    
the non-zero value of $\mbox{Re}[r^{04}_{10}]$ 
indicates a helicity single-flip contribution at the level of a few
percent. The finite negative values of
$r^{04}_{1-1}$ show clear evidence for a helicity-double-flip
contribution.

\begin{figure}[p]
\vfill
\begin{center}
\epsfig{file=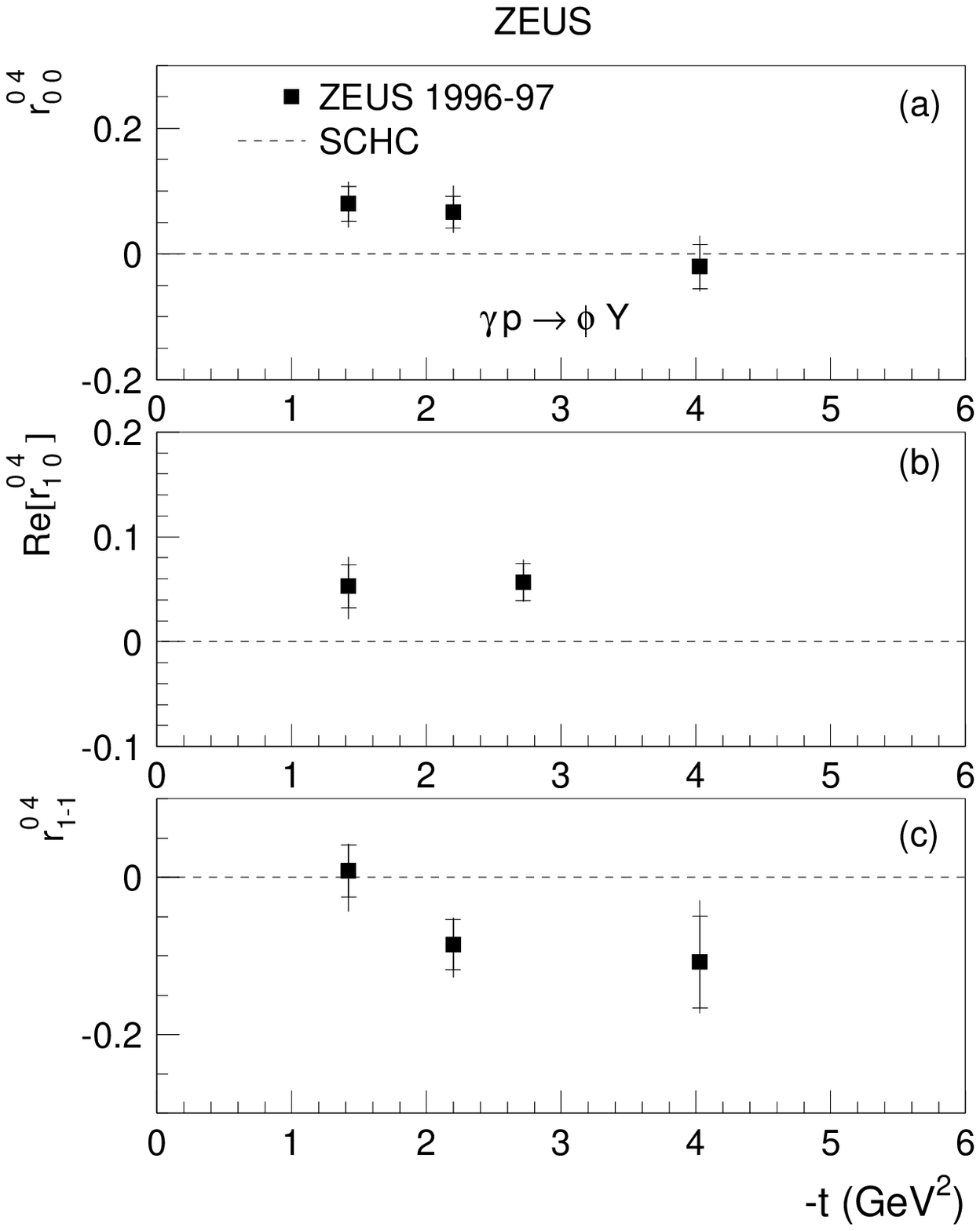,width=1.00\textwidth}
\end{center}  
  \caption{The fitted values of (a) $r^{04}_{00}$,  (b) $\mbox{Re}[r^{04}_{10}]$
and (c) $r^{04}_{1-1}$ for proton-dissociative $\phi$ meson photoproduction 
as a function of $-t$.
  The inner bars indicate
  the statistical uncertainty and the outer bars represent 
  the statistical and
  systematic uncertainties added in quadrature.
The SCHC prediction is shown as the dashed line.
}
\label{fig-phi_hel_res}
\vfill
\end{figure}

\begin{figure}[p]
\vfill
\begin{center}
\epsfig{file=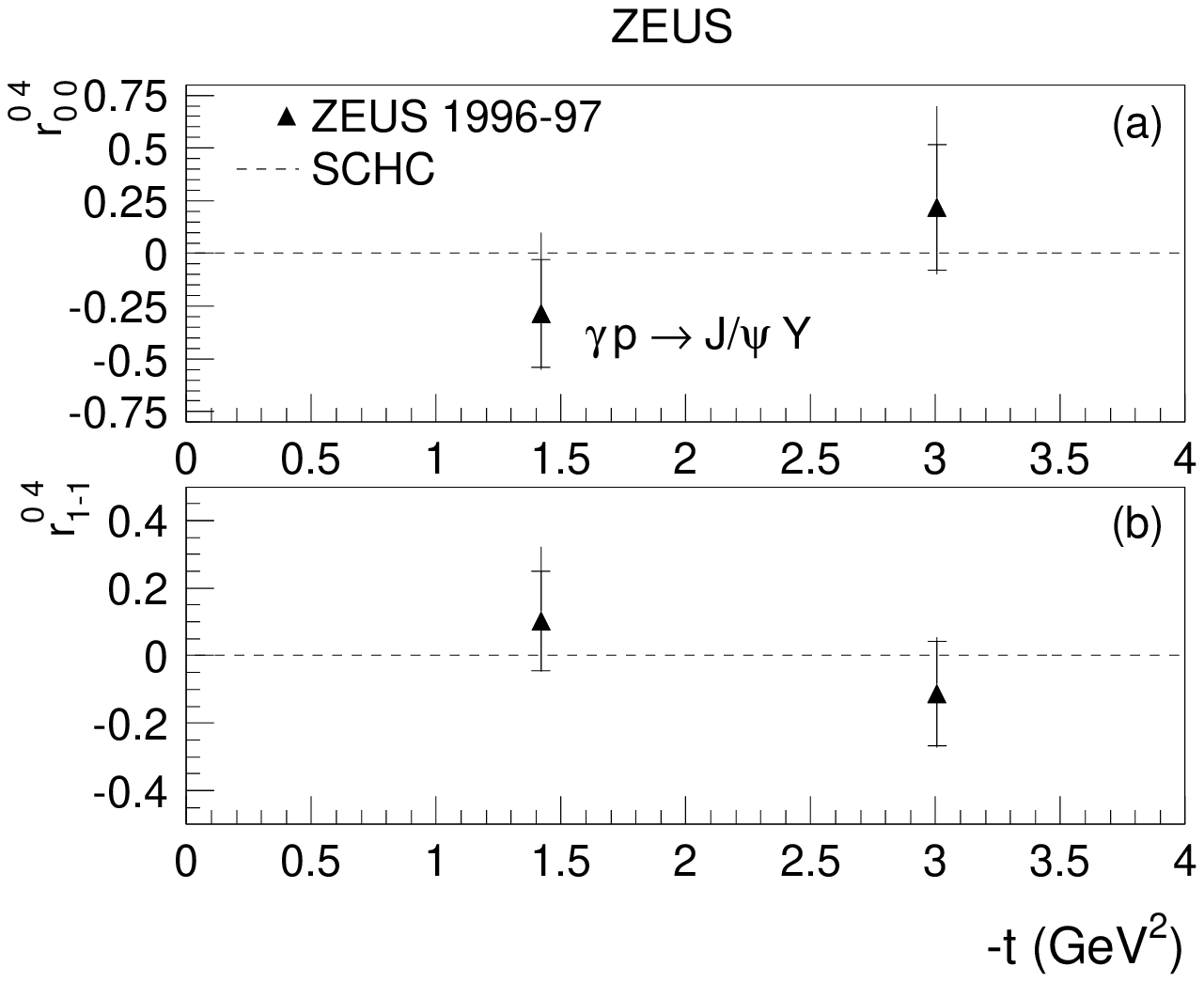,width=1.00\textwidth}
\end{center}  
\caption{The fitted values of (a) $r^{04}_{00}$
and (b) $r^{04}_{1-1}$ for proton-dissociative $J/\psi$ meson photoproduction 
as a function of $-t$.
  The inner bars indicate
  the statistical uncertainty and the outer bars represent 
  the statistical and
  systematic uncertainties added in quadrature.
The SCHC prediction is shown as the dashed line.
}
\label{fig-psi_hel_res}
\vfill
\end{figure}

The measurements for the $\phi$ meson, shown in~\fig{phi_hel_res}, 
display similar features to those for the  $\rho^0$ meson, although
with a smaller statistical significance.
The values of $r^{04}_{00}$
and $r^{04}_{1-1}$ for the $J/\psi$ meson, shown in~\fig{psi_hel_res}, 
are consistent with the SCHC hypothesis.

Although the BFKL calculation of Forshaw and 
Poludniowski~\cite{hep-ph-0107068}
successfully describes the cross sections for all three mesons,
it is not in accord with the observation of SCHC violation. 
This calculation forces SCHC by using a delta-function for the 
meson wave-function.

If a more appropriate meson wave-function for the light mesons is used,
the helicity of the final-state meson will depend 
on the modelling
of the photon fluctuation into a $q\qbar$ pair. 
Within perturbation theory, a photon can 
only split into a $q\qbar$ pair in a chiral-even spin configuration.  
If, in addition, the quark appearing in the $\gamma V$ coupling
is interpreted as a current quark, as assumed by Ginzburg and 
Ivanov~\cite{pr:d54:5523},
real photons will produce light mesons only in the helicity-$0$ 
state because of the chiral nature of the perturbative coupling
in the massless limit. Not only does this model not agree with 
the cross section~\citeHIGHT95, but it is also ruled out by the helicity data. 
Ivanov et al.~\cite{pl:b478:101} have proposed
that a large contribution of light mesons produced in the 
helicity-$\pm 1$ states could arise from $q\qbar$ fluctuations 
in a chiral-odd spin configuration. 
This model fails to describe the cross section and 
does not quantitatively describe the helicity structure.

\section{Summary}
\label{sec-sum}

Diffractive  photoproduction of $\rho^0$, $\phi$ and
$J/\psi$ mesons with proton dissociation has been measured 
in the ZEUS detector at HERA using data
at $W\approx 100 \gev$ for $-t$ 
values up to $12\gev^2$. 
Expressing the differential cross section  
as a power-law, $\diff \sigma_{\gamma p \to V Y}/\diff t \propto (-t)^{-n}$,
yields a value for $n$ that decreases with increasing mass of the meson.
The resulting exponents are: 
$3.21 \pm 0.04$(stat.)$\pm 0.15$(syst.) for $\rho^0$ production; 
$2.7 \pm 0.1$(stat.)$\pm 0.2$(syst.) for $\phi$ production;
and $1.7 \pm 0.2$(stat.)$\pm 0.3$(syst.) for $J/\psi$ production. 
The BFKL calculation of Bartels et al.~\cite{pl:b375:301} 
agrees with the $J/\psi$ data. 
An extension of this model 
by Forshaw and Poludniowski~\cite{hep-ph-0107068}
is able to describe the data for all three mesons.

The cross-section ratios for the $\phi$ and $J/\psi$ with respect
to the $\rho$ as a function of $t$ 
increase with $-t$. In the context of pQCD,
this observation suggests that the production 
mechanism may become flavour independent at larger $-t$ values
than those covered in this analysis. 

The decay-angle analysis for the $\rho^0$ and $\phi$
mesons indicates that the SCHC hypothesis does not hold, since both 
single and double helicity-flip contributions are observed.  
No available pQCD calculation is able to 
describe this result quantitatively.

The effective slope of the Pomeron trajectory, $\alpha^\prime$, 
was determined 
by studying the relative changes of the $W$ dependence of the cross-section
as a function of $t$ for the $\rho^0$ and $\phi$ mesons.
The slopes are
$\alpha^\prime = -0.02 \pm 0.05\mbox{(stat.)} 
_{-0.08}^{+0.04} \mbox{(syst.)} \gev^{-2}$ for the $\rho^0$ meson and
$\alpha^\prime = -0.06 \pm 0.12 \mbox{(stat.)} 
_{-0.09}^{+0.05} \mbox{(syst.)} \gev^{-2}$ for the $\phi$ meson.
These values are lower than the value 
characteristic of low $-t$ hadronic scattering 
($\alpha^\prime=0.25\gev^{-2}$) and lower than those
obtained in elastic vector-meson photoproduction at HERA
at lower $-t$~\cite{epj:c14:213,hep-ex-0201043}.
They are in agreement with pQCD expectations~\cite{pl:b478:101,pl:b375:301} 
and are consistent with the flatness of the Pomeron trajectory
at large $-t$ observed by UA8~\cite{np:b514:3}. 
These results imply that 
large values of $-t$ can provide a suitable hard scale for perturbative
QCD calculations. 

\section{Acknowledgments}
We thank the DESY directorate for their strong support and encouragement.
The remarkable achievements of the HERA machine group were essential for 
the successful completion of this work and are gratefully
acknowledged. We are grateful for the support of the DESY computing and 
network services. The design, construction and installation of the ZEUS 
detector has been made possible by the efforts and ingenuity of many 
people from DESY and the home institutes who are not listed as authors.
We are grateful to J.R. Forshaw and G. Poludniowski for providing 
the results of their calculation and to J.~Bartels and D.Yu~Ivanov 
for useful discussion. 
\vfill\eject


{
\def\bibname{\Large\bf References}
\def\refname{\Large\bf References}
\pagestyle{plain}
\ifzeusbst
  \bibliographystyle{./BiBTeX/bst/l4z_default}
\fi
\ifzdrftbst
  \bibliographystyle{./BiBTeX/bst/l4z_draft}
\fi
\ifzbstepj
  \bibliographystyle{./BiBTeX/bst/l4z_epj}
\fi
\ifzbstnp
  \bibliographystyle{./BiBTeX/bst/l4z_np}
\fi
\ifzbstpl
  \bibliographystyle{./BiBTeX/bst/l4z_pl}
\fi
{\raggedright
\bibliography{./BiBTeX/user/syn.bib,%
              ./BiBTeX/bib/l4z_articles.bib,%
              ./BiBTeX/bib/l4z_books.bib,%
              ./BiBTeX/bib/l4z_conferences.bib,%
              ./BiBTeX/bib/l4z_h1.bib,%
              ./BiBTeX/bib/l4z_misc.bib,%
              ./BiBTeX/bib/l4z_old.bib,%
              ./BiBTeX/bib/l4z_preprints.bib,%
              ./BiBTeX/bib/l4z_replaced.bib,%
              ./BiBTeX/bib/l4z_temporary.bib,%
              ./BiBTeX/bib/l4z_zeus.bib}}

\providecommand{\etal}{et al.\xspace}
\providecommand{\coll}{Coll.\xspace}
\catcode`\@=11
\def\@bibitem#1{%
\ifmc@bstsupport
  \mc@iftail{#1}%
    {;\newline\ignorespaces}%
    {\ifmc@first\else.\fi\orig@bibitem{#1}}
  \mc@firstfalse
\else
  \mc@iftail{#1}%
    {\ignorespaces}%
    {\orig@bibitem{#1}}%
\fi}%
\catcode`\@=12
\begin{mcbibliography}{10}

\bibitem{crittenden:1997:mesons}
J.A.~Crittenden,
\newblock {\em Exclusive Production of Neutral Vector Mesons at the
  Electron-Proton Collider {HERA}},
\newblock Springer Tracts in Modern Physics, Vol. 140.
\newblock Springer, Berlin, Germany, 1997\relax
\relax
\bibitem{rmp:71:1275}
H.~Abramowicz and A.~Caldwell,
\newblock Rev.\ Mod.\ Phys.{} {\bf 71},~1275~(1999)\relax
\relax
\bibitem{np:b231:189}
A.~Donnachie and P.V.~Landshoff,
\newblock Nucl.\ Phys.{} {\bf B~231},~189~(1984)\relax
\relax
\bibitem{pl:b395:311}
J.R.~Cudell, K.~Kang and S.K.~Kim,
\newblock Phys.\ Lett.{} {\bf B~395},~311~(1997)\relax
\relax
\bibitem{collins:1977:regge}
P.D.B.~Collins,
\newblock {\em An Introduction to {Regge} Theory and High Energy Physics}.
\newblock Cambridge University Press, 1977\relax
\relax
\bibitem{pl:b470:243}
A.~Donnachie and P.V.~Landshoff,
\newblock Phys.\ Lett.{} {\bf B~470},~243~(1999)\relax
\relax
\bibitem{pl:b478:146}
A.~Donnachie and P.V.~Landshoff,
\newblock Phys.\ Lett.{} {\bf B~478},~146~(2000)\relax
\relax
\bibitem{hep-ph-0112242}
E.~Martynov, E.~Predazzi and A.~Prokudin,
\newblock Preprint \mbox{hep-ph/0112242}, 2001.
\newblock Subm.\ to Eur.~Phys.~J\relax
\relax
\bibitem{anphy:11:1}
J.J.~Sakurai,
\newblock Ann.~Phys.{} {\bf 11},~1~(1960)\relax
\relax
\bibitem{prl:22:981}
J.J.~Sakurai,
\newblock Phys.\ Rev.\ Lett.{} {\bf 22},~981~(1969)\relax
\relax
\bibitem{zfp:c57:89}
M.G.~Ryskin,
\newblock Z.\ Phys.{} {\bf C~57},~89~(1993)\relax
\relax
\bibitem{pr:d50:3134}
S.J.~Brodsky \etal,
\newblock Phys.\ Rev.{} {\bf D~50},~3134~(1994)\relax
\relax
\bibitem{prl:63:1914}
L.~Frankfurt and M.~Strikman,
\newblock Phys.\ Rev.\ Lett.{} {\bf 63},~1914~(1989)\relax
\relax
\bibitem{pl:b284:123}
A.H.~Mueller and W.K.~Tang,
\newblock Phys.\ Lett.{} {\bf B~284},~123~(1992)\relax
\relax
\bibitem{epj:c14:213}
ZEUS \coll, J.~Breitweg \etal,
\newblock Eur.\ Phys.\ J.{} {\bf C~14},~213~(2000)\relax
\relax
\bibitem{hep-ex-0201043}
ZEUS \coll, S.~Chekanov \etal,
\newblock Preprint \mbox{DESY-02-008} (\mbox{hep-ex-0201043}), DESY, 2002.
\newblock Subm.\ to Eur.~Phys.~J\relax
\relax
\bibitem{npps:99a:24}
N.I.~Kochelev \etal,
\newblock Nucl.\ Phys.\ Proc.\ Suppl.{} {\bf 99~A},~24~(2001)\relax
\relax
\bibitem{zfp:c68:137}
J.R.~Forshaw and M.G.~Ryskin,
\newblock Z.\ Phys.{} {\bf C~68},~137~(1995)\relax
\relax
\bibitem{pr:d53:3564}
D.Yu.~Ivanov,
\newblock Phys.\ Rev.{} {\bf D~53},~3564~(1996)\relax
\relax
\bibitem{pr:d54:5523}
I.F.~Ginzburg and D.Yu.~Ivanov,
\newblock Phys.\ Rev.{} {\bf D~54},~5523~(1996)\relax
\relax
\bibitem{pl:b375:301}
J.~Bartels \etal,
\newblock Phys.\ Lett.{} {\bf B~375},~301~(1996)\relax
\relax
\bibitem{pl:b478:101}
D.Yu.~Ivanov \etal,
\newblock Phys.\ Lett.{} {\bf B~478},~101~(2000).
\newblock Erratum in Phys.~Lett.~{\bf B~498},~295~(2001)\relax
\relax
\bibitem{hep-ph-0107068}
J.R~Forshaw and G.~Poludniowski,
\newblock Preprint \mbox{hep-ph/0107068}, 2001\relax
\relax
\bibitem{epj:c5:461}
M.~Gl\"uck, E.~Reya and A.~Vogt,
\newblock Eur.\ Phys.\ J.{} {\bf C~5},~461~(1998)\relax
\relax
\bibitem{jetp:45:199}
E.A.~Kuraev, L.N.~Lipatov and V.S.~Fadin,
\newblock Sov.\ Phys.\ JETP{} {\bf 45},~199~(1977)\relax
\relax
\bibitem{sovjnp:28:822}
Ya.Ya.~Balitski\u i and L.N.~Lipatov,
\newblock Sov.\ J.\ Nucl.\ Phys.{} {\bf 28},~822~(1978)\relax
\relax
\bibitem{pl:b31:387}
F.J.~Gilman \etal,
\newblock Phys.\ Lett.{} {\bf B~31},~387~(1970)\relax
\relax
\bibitem{epj:c15:1}
Particle Data Group, D.E. Groom \etal,
\newblock Eur.\ Phys.\ J.{} {\bf C~15},~1~(2000)\relax
\relax
\bibitem{zeus:1993:bluebook}
ZEUS \coll, U.~Holm~(ed.),
\newblock {\em The {ZEUS} Detector}.
\newblock Status Report (unpublished), DESY (1993),
\newblock available on
  \texttt{http://www-zeus.desy.de/bluebook/bluebook.html}\relax
\relax
\bibitem{nim:a279:290}
N.~Harnew \etal,
\newblock Nucl.\ Inst.\ Meth.{} {\bf A~279},~290~(1989)\relax
\relax
\bibitem{npps:b32:181}
B.~Foster \etal,
\newblock Nucl.\ Phys.\ Proc.\ Suppl.{} {\bf B~32},~181~(1993)\relax
\relax
\bibitem{nim:a338:254}
B.~Foster \etal,
\newblock Nucl.\ Inst.\ Meth.{} {\bf A~338},~254~(1994)\relax
\relax
\bibitem{nim:a309:77}
M.~Derrick \etal,
\newblock Nucl.\ Inst.\ Meth.{} {\bf A~309},~77~(1991)\relax
\relax
\bibitem{nim:a309:101}
A.~Andresen \etal,
\newblock Nucl.\ Inst.\ Meth.{} {\bf A~309},~101~(1991)\relax
\relax
\bibitem{nim:a321:356}
A.~Caldwell \etal,
\newblock Nucl.\ Inst.\ Meth.{} {\bf A~321},~356~(1992)\relax
\relax
\bibitem{nim:a336:23}
A.~Bernstein \etal,
\newblock Nucl.\ Inst.\ Meth.{} {\bf A~336},~23~(1993)\relax
\relax
\bibitem{zfp:c75:421}
ZEUS \coll, J.~Breitweg \etal,
\newblock Z.\ Phys.{} {\bf C~75},~421~(1997)\relax
\relax
\bibitem{acpp:b32:2025}
J.~Andruszk\'ow \etal,
\newblock Acta Phys.\ Pol.{} {\bf B~32},~2025~(2001)\relax
\relax
\bibitem{np:b61:381}
K.~Schilling and G.~Wolf,
\newblock Nucl.\ Phys.{} {\bf B~61},~381~(1973)\relax
\relax
\bibitem{thesis:kasprzak:1994}
M.~Kasprzak,
\newblock {\em Inclusive Properties of Diffractive and Non-diffractive
  Photoproduction at {HERA}}.
\newblock Ph.D.\ Thesis, Warsaw University, Warsaw, Poland, Report \mbox{DESY
  F35D-96-16}, DESY, 1996\relax
\relax
\bibitem{thesis:adamczyk:1999}
L.~Adamczyk,
\newblock {\em Vector Meson Photoproduction at Large Momentum Transfer at
  {HERA}}.
\newblock Ph.D.\ Thesis, University of Mining and Metallurgy, Cracow, Poland,
  Report \mbox{DESY-THESIS-1999-045}, DESY, 1999\relax
\relax
\bibitem{proc:mc:1998:396}
B.~List and A.~Mastroberardino,
\newblock {\em Proc.\ Workshop on Monte Carlo Generators for {HERA} Physics},
  p.~396.
\newblock DESY, Hamburg, Germany (1999).
\newblock Also in preprint \mbox{DESY-PROC-1999-02},
\newblock available on \texttt{http://www.desy.de/\til heramc/}\relax
\relax
\bibitem{prl:48:1451}
R.L.~Cool \etal,
\newblock Phys.\ Rev.\ Lett.{} {\bf 48},~1451~(1982)\relax
\relax
\bibitem{rncim:6:1}
R.~Hagedorn,
\newblock Riv.~Nuovo~Cim.{} {\bf 6},~1~(1984)\relax
\relax
\bibitem{cpc:82:74}
T.~Sj\"ostrand,
\newblock Comp.\ Phys.\ Comm.{} {\bf 82},~74~(1994)\relax
\relax
\bibitem{tech:cern-dd-ee-84-1}
R.~Brun et al.,
\newblock {\em {\sc geant3}},
\newblock Technical Report CERN-DD/EE/84-1, CERN, 1987\relax
\relax
\bibitem{thesis:klimek:2001}
K.H.~Klimek,
\newblock {\em Cross Section Measurement of Vector Meson Quasi-Photoproduction
  at High Four-Momentum Transfer Using the ZEUS Detector at the HERA Collider}.
\newblock Ph.D. Thesis, Institute of Nuclear Physics, Cracow, Poland, Report
  \mbox{DESY-THESIS-2001-053}, DESY, 2001\relax
\relax
\bibitem{prep:15:181}
V.M.~Budnev \etal,
\newblock Phys.\ Rep.{} {\bf 15C},~181~(1974)\relax
\relax
\bibitem{np:b77:240}
D.P.~Roy and R.G.~Roberts,
\newblock Nucl.\ Phys.{} {\bf B~77},~240~(1974)\relax
\relax
\bibitem{pl:19:702}
P.~S\"oding,
\newblock Phys.\ Lett.{} {\bf 19},~702~(1966)\relax
\relax
\bibitem{ncim:34:1644}
J.D.~Jackson,
\newblock Nuovo Cimento{} {\bf 34},~1644~(1964)\relax
\relax
\bibitem{shep:11:51}
H.~Abramowicz, L.~Frankfurt and M.~Strikman,
\newblock Surv.\ High Energy Phys.{} {\bf 11},~51~(1997)\relax
\relax
\bibitem{pl:b483:360}
H1 \coll, C.~Adloff \etal,
\newblock Phys.\ Lett.{} {\bf B~483},~360~(2000)\relax
\relax
\bibitem{pl:b487:273}
ZEUS \coll, J.~Breitweg \etal,
\newblock Phys.\ Lett.{} {\bf B~487},~273~(2000)\relax
\relax
\bibitem{epj:c6:603}
ZEUS \coll, J.~Breitweg \etal,
\newblock Eur.\ Phys.\ J.{} {\bf C~6},~603~(1999)\relax
\relax
\bibitem{epj:c13:371}
H1 \coll, C.~Adloff \etal,
\newblock Eur.\ Phys.\ J.{} {\bf C~13},~371~(2000)\relax
\relax
\bibitem{epj:c10:373}
H1 \coll, C.~Adloff \etal,
\newblock Eur.\ Phys.\ J.{} {\bf C~10},~373~(1999)\relax
\relax
\bibitem{priv:meyer:2002}
A.B.~Meyer, private communication\relax
\relax
\bibitem{np:b514:3}
UA8 \coll, A.~Brandt \etal,
\newblock Nucl.\ Phys.{} {\bf B~514},~3~(1998)\relax
\relax
\bibitem{thesis:kowal:2002}
A.~Kowal,
\newblock {\em Helicity Analysis of Vector Mesons Produced in
  Proton-dissociative Diffractive Photoproduction at Large Momentum Transfer at
  HERA}.
\newblock Ph.D. Thesis, University of Mining and Metallurgy, Cracow, Poland,
  2002.
\newblock Unpublished\relax
\relax
\bibitem{pl:b377:259}
ZEUS \coll, M.~Derrick \etal,
\newblock Phys.\ Lett.{} {\bf B~377},~259~(1996)\relax
\relax
\bibitem{pl:b380:220}
ZEUS \coll, M.~Derrick \etal,
\newblock Phys.\ Lett.{} {\bf B~380},~220~(1996)\relax
\relax
\bibitem{zfp:c75:215}
ZEUS \coll, J.~Breitweg \etal,
\newblock Z.\ Phys.{} {\bf C~75},~215~(1997)\relax
\relax
\end{mcbibliography}
}
\vfill\eject

\begin{table}[ht!] 

\begin{center}
\begin{tabular}{|c|c|c|}
\hline
\multicolumn{3}{|c|}{ZEUS 1996-97 $\gamma p \to \rho^0 Y$} \\ \hline 
$-t$ range & $\langle -t \rangle$ & \multicolumn{1}{|c|}{$\diff\sigma/\diff|t|$} \\ 
 $(\gev^2)$ & $(\gev^2)$ & \multicolumn{1}{|c|}{$(nb/\gev^2)$} \\ \hline

1.20-1.50 & 1.33   &  163   $\pm$    6   $_{-19}    ^{+14}$   $_{-12}    ^{+7} $ \\ \hline
1.50-2.00 & 1.71   &   80.1  $\pm$    2.5   $_{- 8.5}    ^{+ 5.5}$   $_{- 8.5}    ^{+ 2.0} $ \\ \hline
2.00-2.50 & 2.22   &   36.9  $\pm$    1.4   $_{- 2.8}    ^{+ 3.4}$   $_{- 2.6}    ^{+ 0.4} $ \\ \hline
2.50-3.00 & 2.72   &   17.5  $\pm$    0.9   $_{- 1.5}    ^{+ 1.3}$   $_{- 1.7}    ^{+ 0.1} $ \\ \hline
3.00-3.50 & 3.23   &   10.7  $\pm$    0.7   $_{- 0.7}    ^{+ 0.8}$   $_{- 1.2}    ^{+ 0.3} $ \\ \hline
3.50-4.00 & 3.73   &    6.98  $\pm$    0.59   $_{- 0.41}    ^{+ 0.81}$   $_{- 0.97}    ^{+ 0.39} $ \\ \hline
4.00-5.00 & 4.43   &    3.39  $\pm$    0.27   $_{- 0.50}    ^{+ 0.78}$   $_{- 0.57}    ^{+ 0.23} $ \\ \hline
5.00-6.85 & 5.75   &    1.41  $\pm$    0.15   $_{- 0.49}    ^{+ 0.35}$   $_{- 0.25}    ^{+ 0.20} $ \\ \hline
6.85-10.0 & 8.04   &    0.46  $\pm$    0.07   $_{- 0.22}    ^{+ 0.17}$   $_{- 0.13}    ^{+ 0.08} $ \\  
\hline 
 
\end{tabular}
\end{center}

\caption{The differential cross section, 
$\diff\sigma_{\gamma p \to \rho^0 Y}/\diff |t|$, 
for $x>0.01$ and $80<W<120 \gev$. 
The first uncertainty is the statistical, the second the systematic and
the last is due to the modelling of the proton-dissociation process.
The normalisation uncertainty of 10\% is not included.
\label{tab-rho_cros}}
\end{table}

\begin{table}[ht!]

\begin{center}
\begin{tabular}{|c|c|c|}
\hline
\multicolumn{3}{|c|}{ZEUS 1996-97 $\gamma p \to \phi Y$} \\ \hline 
$-t$ range & $\langle -t \rangle$ & $\diff\sigma/\diff|t|$) \\ 
 $(\gev^2)$ & $(\gev^2)$ & $(nb/\gev^2)$ \\ \hline 
 
 1.20-1.55 &  1.36  &  19.2   $\pm$   1.1    $_{- 2.1}    ^{+ 2.5}$   $_{- 1.7}    ^{+ 1.0}$\\\hline
 1.55-1.90 &  1.71  &  12.4   $\pm$   0.7    $_{- 1.1}    ^{+ 1.8}$   $_{- 0.9}    ^{+ 0.3}$\\\hline
 1.90-2.30 &  2.08  &   7.43   $\pm$   0.53    $_{- 0.72}    ^{+ 0.68}$   $_{- 0.41}    ^{+ 0.30}$\\\hline
 2.30-2.70 &  2.48  &   4.63   $\pm$   0.40    $_{- 0.43}    ^{+ 0.49}$   $_{- 0.08}    ^{+ 0.17}$\\\hline
 2.70-3.20 &  2.93  &   2.55   $\pm$   0.26    $_{- 0.29}    ^{+ 0.30}$   $_{- 0.08}    ^{+ 0.15}$\\\hline
 3.20-3.80 &  3.47  &   1.55   $\pm$   0.20    $_{- 0.25}    ^{+ 0.33}$   $_{- 0.07}    ^{+ 0.13}$\\\hline
 3.80-4.80 &  4.24  &   0.89   $\pm$   0.18    $_{- 0.14}    ^{+ 0.31}$   $_{- 0.07}    ^{+ 0.13}$\\\hline
 4.80-6.50 &  5.53  &   0.36   $\pm$   0.08    $_{- 0.12}    ^{+ 0.09}$   $_{- 0.06}    ^{+ 0.10}$\\\hline
 
\end{tabular}
\end{center}

\caption{The differential cross section, 
$\diff\sigma_{\gamma p \to \phi Y}/\diff |t|$, 
for $x>0.01$ and $80<W<120 \gev$. 
The first uncertainty is the statistical, the second the systematic and
the last is due to the modelling of the proton-dissociation process.
The normalisation uncertainty of 10\% is not included.
\label{tab-phi_cros}}
\end{table}

\begin{table}[ht!]

\begin{center}
\begin{tabular}{|c|c|c|}
\hline
\multicolumn{3}{|c|}{ZEUS 1996-97 $\gamma p \to J/\psi Y$} \\ \hline 
$-t$ range & $\langle -t \rangle$ & $\diff\sigma/\diff|t|$ \\ 
 $(\gev^2)$ & $(\gev^2)$ & $(nb/\gev^2)$ \\ \hline 
 
1.20-1.50  & 1.34    &16.0   $\pm$  2.4   $_{-  3.4}     ^{+1.6}$    $_{- 4.8}    ^{+ 1.6}$ \\\hline
1.50-2.00  & 1.74    & 9.5   $\pm$  1.5   $_{-  1.4}     ^{+1.4}$    $_{- 1.2}    ^{+ 0.3}$ \\\hline
2.00-2.50  & 2.24    & 8.3   $\pm$  1.5   $_{-  1.6}     ^{+1.6}$    $_{- 2.4}    ^{+ 0.3}$ \\\hline
2.50-3.25  & 2.84    & 4.98   $\pm$  0.95   $_{-  0.90}     ^{+0.89}$    $_{- 0.63}    ^{+ 0.38}$ \\\hline
3.25-4.50  & 3.81    & 2.5   $\pm$  0.6   $_{-  1.0}     ^{+0.8}$    $_{- 0.4}    ^{+ 0.4}$ \\\hline
4.50-6.50  & 5.37    & 1.33   $\pm$  0.45   $_{-  0.41}     ^{+0.80}$    $_{- 0.39}    ^{+ 0.16}$ \\\hline
 
\end{tabular}
\end{center}

\caption{The differential cross section, 
$\diff\sigma_{\gamma p \to J/\psi Y}/\diff |t|$, 
for $x>0.01$ and $80<W<120 \gev$. 
The first uncertainty is the statistical, the second the systematic and
the last is due to the modelling of the proton-dissociation process.
The normalisation uncertainty of 10\% is not included.
\label{tab-psi_cros}}
\end{table}

\begin{table}[ht!]

\begin{center}
\begin{tabular}{|c|c|c|}
\hline
\multicolumn{3}{|c|}{ZEUS 1996-97 } \\ \hline 
$-t$ range & $\langle -t \rangle$ & $\frac{\diff\sigma}{\diff t}(\gamma p \rightarrow \phi Y) / \frac{\diff\sigma}{\diff t}(\gamma p \rightarrow \rho^0 Y$) \\ \hline 
 
1.20-1.55   & 1.35   &   0.124 $\pm$    0.008    $_{- 0.012}   ^{+  0.020}$    $_{- 0.013}   ^{+  0.014}$ \\\hline
1.55-1.90   & 1.71   &   0.156 $\pm$    0.011    $_{- 0.010}   ^{+  0.023}$    $_{- 0.013}   ^{+  0.014}$ \\\hline
1.90-2.30   & 2.08   &   0.176 $\pm$    0.015    $_{- 0.017}   ^{+  0.021}$    $_{- 0.010}   ^{+  0.011}$ \\\hline
2.30-2.70   & 2.48   &   0.167 $\pm$    0.016    $_{- 0.013}   ^{+  0.013}$    $_{- 0.006}   ^{+  0.006}$ \\\hline
2.70-3.20   & 2.93   &   0.185 $\pm$    0.021    $_{- 0.034}   ^{+  0.014}$    $_{- 0.013}   ^{+  0.013}$ \\\hline
3.20-3.80   & 3.47   &   0.189 $\pm$    0.027    $_{- 0.038}   ^{+  0.020}$    $_{- 0.020}   ^{+  0.021}$ \\\hline
3.80-4.80   & 4.23   &   0.223 $\pm$    0.047    $_{- 0.042}   ^{+  0.057}$    $_{- 0.033}   ^{+  0.034}$ \\\hline
4.80-6.50   & 5.49   &   0.225 $\pm$    0.052    $_{- 0.060}   ^{+  0.055}$    $_{- 0.052}   ^{+  0.055}$ \\\hline

\end{tabular}
\end{center}

\caption{The ratio of the differential cross sections, 
$\diff\sigma/\diff t$, for $\phi$ to $\rho^0$ proton-dissociative 
photoproduction for $x>0.01$ and $80<W<120 \gev$.
The first uncertainty is the statistical, the second the systematic and
the last is due to the modelling of the proton-dissociation process.  
\label{tab-phi_rho}}
\end{table}

\begin{table}[ht!]

\begin{center}
\begin{tabular}{|c|c|c|}
\hline
\multicolumn{3}{|c|}{ZEUS 1996-97 } \\ \hline 
$-t$ range & $\langle -t \rangle$ & $\frac{\diff\sigma}{\diff t}(\gamma p \rightarrow J/\psi Y) / \frac{\diff\sigma}{\diff t}(\gamma p \rightarrow \rho^0 Y$) \\ \hline 

1.20-1.50 &   1.33   &  0.099  $\pm$    0.015    $_{- 0.019}^{+     0.021}$    $_{- 0.007}^{+     0.010}$ \\\hline
1.50-2.00 &   1.71   &  0.119  $\pm$    0.019    $_{- 0.017}^{+     0.027 }$   $_{- 0.007}^{+     0.007}$ \\\hline
2.00-2.50 &   2.22   &  0.226  $\pm$    0.042    $_{- 0.037}^{+     0.056  }$  $_{- 0.008 }^{+    0.005}$ \\\hline
2.50-3.25 &   2.82   &  0.325  $\pm$    0.063    $_{- 0.059}^{+     0.020  }$  $_{- 0.018 }^{+    0.013}$ \\\hline
3.25-4.50 &   3.75   &  0.409  $\pm$    0.097    $_{- 0.121}^{+     0.078 }$   $_{- 0.046 }^{+    0.038}$ \\\hline
4.50-6.50 &   5.28   &  0.72  $\pm$    0.25    $_{- 0.22}^{+     0.20 }$   $_{- 0.15 }^{+    0.11}$ \\\hline

\end{tabular}
\end{center}

\caption{The ratio of the differential cross sections, 
$\diff\sigma/\diff t$, for $J/\psi$ to $\rho^0$ proton-dissociative 
photoproduction 
for $x>0.01$ and $80<W<120 \gev$. 
The first uncertainty is the statistical, the second the systematic and
the last is due to the modelling of the proton-dissociation process.
\label{tab-psi_rho}}
\end{table}

\begin{table}[ht!]
\centering
\begin{tabular}{|c|c|c|c|c|}
\hline
\multicolumn{5}{|c|}{ZEUS 1996-97 $\gamma p \to \rho^0 Y$} \\\hline
$-t$ range & $\langle -t \rangle$ & $r^{04}_{00}$ & $Re\left[r^{04}_{10}\right]$ & $r^{04}_{1-1}$ \\
(GeV$^2$) & (GeV$^2$) & & &
\\\hline
1.1--1.7 & 1.34 & 0.022$\pm$0.015$^{+0.016}_{-0.018}$ &0.054$\pm$0.012$^{+0.017}_{-0.010}$ & -0.143$\pm$0.014$^{+0.023}_{-0.031}$  
\\\hline
1.7--2.9 & 2.14 & 0.042$\pm$0.017$^{+0.018}_{-0.021}$ & 0.045$\pm$0.013$^{+0.016}_{-0.010}$& -0.158$\pm$0.017$^{+0.020}_{-0.022}$ 
\\\hline
2.9--4.0 & 3.35 & 0.037$\pm$0.034$^{+0.032}_{-0.036}$ & & -0.129$\pm$0.032$^{+0.023}_{-0.050}$ 
\\\hline
4.0--12 & 5.67 & 0.090$\pm$0.049$^{+0.035}_{-0.040}$ & & -0.252$\pm$0.047$^{+0.044}_{-0.023}$ 
\\\hline
2.9--12 & 4.38 & & 0.047$\pm$0.020$^{+0.009}_{-0.017}$ &
\\\hline
\end{tabular}
\caption{The spin density matrix elements for $\gamma p \to \rho^0 Y$
for $80<W<120 \gev$. 
The first uncertainty is the statistical and the second the systematic.
\label{tab-rhoSDM}}
\end{table}

\begin{table}[ht!]
\centering
\begin{tabular}{|c|c|c|c|c|}
\hline
\multicolumn{5}{|c|}{ZEUS 1996-97 $\gamma p \to \phi Y$} \\\hline
$-t$ range & $\langle -t \rangle$ & $r^{04}_{00}$ & $Re\left[r^{04}_{10}\right]$ & $r^{04}_{1-1}$ \\
(GeV$^2$) & (GeV$^2$) & & &
\\\hline
1.2--1.7 & 1.42 & 0.080$\pm$0.028$^{+0.020}_{-0.026}$ & 0.053$\pm$0.021$^{+0.019}_{-0.024}$&  0.008$\pm$0.033$^{+0.011}_{-0.039}$  
\\\hline
1.7--3.0 & 2.20 & 0.066$\pm$0.025$^{+0.034}_{-0.021}$ & & -0.085$\pm$0.032$^{+0.023}_{-0.028}$ 
\\\hline
3.0--7.0 & 4.03 & -0.020$\pm$0.035$^{+0.034}_{-0.018}$ & & -0.108$\pm$0.058$^{+0.053}_{-0.030}$
\\\hline
1.7--7.0 & 2.72 & & 0.057$\pm$0.018$^{+0.013}_{-0.007}$ &
\\\hline
\end{tabular}
\caption{The spin density matrix elements for $\gamma p \to \phi Y$
for $80<W<120 \gev$.
The first uncertainty is the statistical and the second the systematic.
\label{tab-phiSDM}}
\end{table}

\begin{table}[ht!]
\centering
\begin{tabular}{|c|c|c|c|}
\hline
\multicolumn{4}{|c|}{ZEUS 1996-97 $\gamma p \to J/\psi Y$} \\\hline
$-t$ range & $\langle -t \rangle$ & $r^{04}_{00}$ & $r^{04}_{1-1}$ \\
(GeV$^2$) & (GeV$^2$) & & 
\\\hline
1.1--1.8 & 1.42 & -0.28$\pm$0.26$^{+0.28}_{-0.08}$ &  0.10$\pm$0.15$^{+0.16}_{-0.03}$ 
\\\hline
1.8--6.5 & 3.01 &  0.22$\pm$0.30$^{+0.38}_{-0.11}$ & -0.11$\pm$0.16$^{+0.06}_{-0.04}$
\\\hline
\end{tabular}
\caption{The spin density matrix elements for $\gamma p \to J/\psi Y$
for $80<W<120 \gev$. 
The first uncertainty is the statistical and the second the systematic.
\label{tab-jpsiSDM}}
\end{table}

\clearpage

\clearpage

\begin{figure}[p]
\vfill
\begin{center}
\epsfig{file=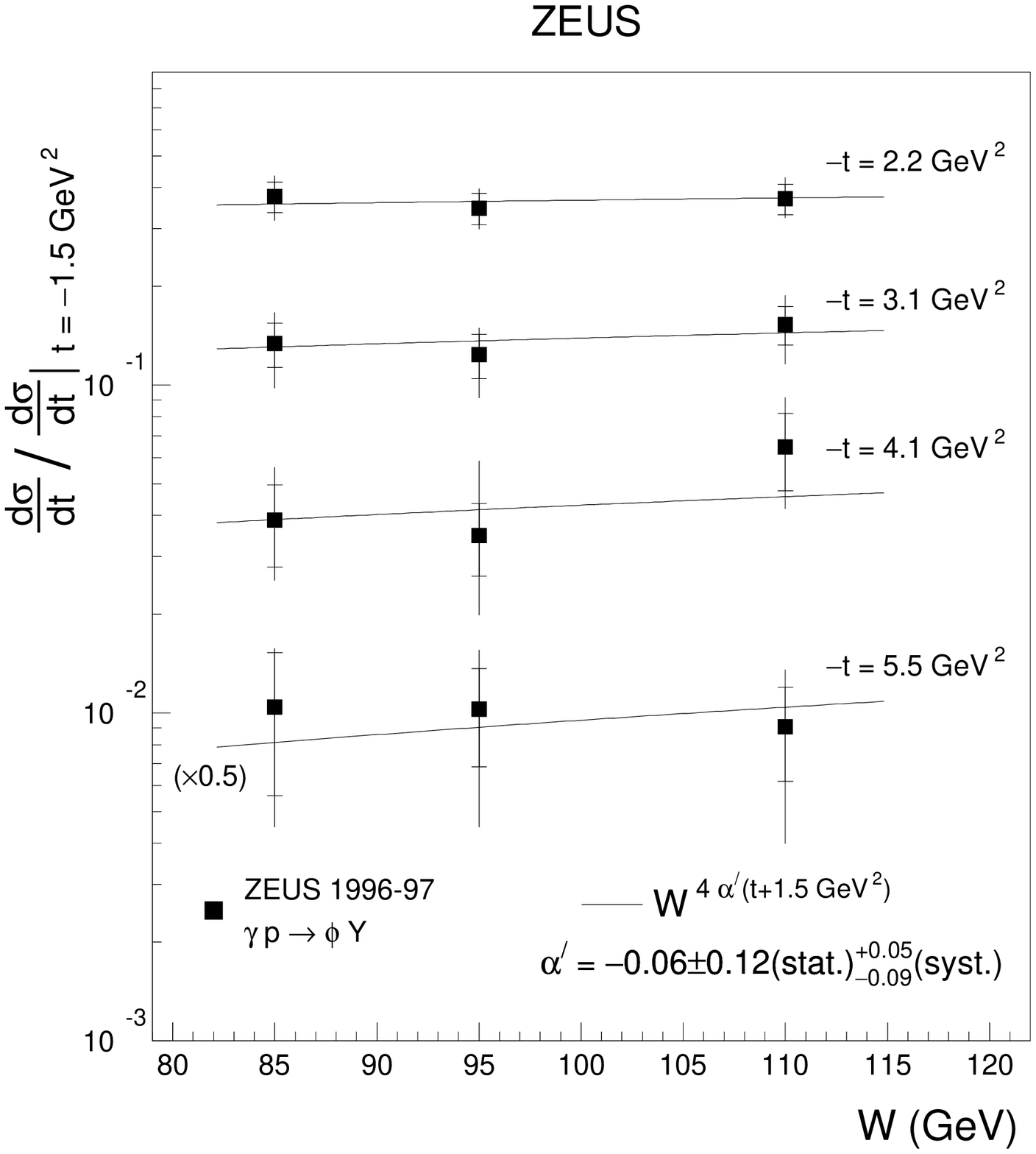,width=1.00\textwidth}
\end{center}  
  \caption{
The ratios of $\phi$ production cross sections 
$\frac{\diff \sigma}{\diff t} / \frac{\diff \sigma}{\diff t}|_{t=t_0}$,
for $-t_0=1.5$ {\rm GeV}$^2$, as a function of $W$ in four $t$ intervals.
The lines represent the result of the fit
with Eq.~\eq{alphap}.
  The inner bars indicate
  the statistical uncertainty and the outer bars represent 
  the statistical and
  systematic uncertainties added in quadrature.
}
  \label{fig-phi_delta}
\vfill
\end{figure}
\clearpage

\clearpage

\clearpage

\clearpage


%
%
\end{document}